\documentclass[12pt]{article}

\usepackage{amsmath}
\usepackage{amsfonts,amssymb,latexsym}
\usepackage[nosort]{cite}
\usepackage[hyperref,bulletsep]{collect}
\def\baselinestretch{1.1}
\newcommand{\pls}{\!+\!}

\newcommand{\mathon}{\mathversion{bold}}
\newcommand{\mathoff}{\mathversion{normal}}

\newcommand{\Ref}[1]{(\ref{#1})}

\newcommand{\ie}{{\it i.e.~}}
\newcommand{\eg}{{\it e.g.~}}

\newcommand{\bea}{\begin{eqnarray}} \newcommand{\eea}{\end{eqnarray}}
\newcommand{\ben}{\begin{displaymath}\nn} \newcommand{\een}{\end{displaymath}}
 
\newcommand{\nn}{\nonumber} \newcommand{\non}{\nonumber\\}

\makeatletter
\let\old@makecaption=\@makecaption
\def\@makecaption{\small\old@makecaption}
\makeatother

\makeatletter
\@addtoreset{equation}{section}
\makeatother

\makeatletter
\let\old@startsection=\@startsection
\renewcommand{\@startsection}[6]{\old@startsection{#1}{#2}{#3}{#4}{#5}{#6\mathversion{bold}}}
\makeatother

\let\oldPhi=\Phi
\let\oldPsi=\Psi
\let\oldGamma=\Gamma
\let\oldSigma=\Sigma
\renewcommand{\Phi}{\mathnormal{\oldPhi}}
\renewcommand{\Psi}{\mathnormal{\oldPsi}}
\renewcommand{\Gamma}{\mathnormal{\oldGamma}}
\renewcommand{\Sigma}{\mathnormal{\oldSigma}}

\newcommand{\sfrac}[2]{{\textstyle\frac{#1}{#2}}}
\newcommand{\half}{\sfrac{1}{2}}

\newcommand{\superN}{\mathcal{N}}
\newcommand{\gym}{g_{\scriptscriptstyle\mathrm{YM}}}
\newcommand{\Tr}{\mathop{\mathrm{Tr}}}
\newcommand{\sign}{\mathop{\mathrm{sign}}}
\newcommand{\indup}[1]{_{\mathrm{#1}}}

\newcommand{\lrbrk}[1]{\left(#1\right)}
\newcommand{\bigbrk}[1]{\bigl(#1\bigr)}

\newcommand{\nln}{\nonumber\\}
\newcommand{\nl}{\nonumber\\&&\mathord{}}

\newcommand{\earel}[1]{\mathrel{}&#1&\mathrel{}}
\newcommand{\eq}{\earel{=}}
\newenvironment{myeqnarray}{\arraycolsep0pt\begin{eqnarray}}{\end{eqnarray}\ignorespacesafterend}
\newenvironment{myeqnarray*}{\arraycolsep0pt\begin{eqnarray*}}{\end{eqnarray*}\ignorespacesafterend}

\def\[{\begin{equation}}
\def\]{\end{equation}}
\def\<{\begin{myeqnarray}}
\def\>{\end{myeqnarray}}

\ifx\href\asklfhas\newcommand{\href}[2]{#2}\fi
\newcommand{\arxivno}[1]{\href{http://arxiv.org/abs/#1}{#1}}

\newcommand{\reps}{\mathcal{R}}
\newcommand{\yfat}{\mathbf{y}}
\newcommand{\wfat}{\mathbf{w}}
\newcommand{\tlong}{\mathcal{T}\indup{sconf}}
\newcommand{\tkk}{\mathcal{T}\indup{KK}}
\newcommand{\tten}{\mathcal{T}\indup{SO(10,2)}}
\newcommand{\zbps}{\mathcal{Z}\indup{BPS}}
\newcommand{\zsym}{\mathcal{Z}\indup{SYM}}
\newcommand{\zten}{\mathcal{Z}\indup{SO(10,2)}}
\newcommand{\zlong}{\mathcal{Z}\indup{long}}
\newcommand{\zprim}{\mathcal{Z}\indup{sconf}}

\begin{document}

\thispagestyle{empty}

\begin{center}

{\small\ttfamily AEI~2003-084\hspace*{0.5cm} 
ROM2F/03/28\hspace*{0.5cm} 
ITP-UU-03/56\hspace*{0.5cm}
SPIN-03/37\hspace*{0.5cm}
\arxivno{hep-th/0310292}}

\end{center}

\bigskip

\begin{center}

\renewcommand{\thefootnote}{\fnsymbol{footnote}}

{\mathon\bf\Large On the spectrum of AdS/CFT beyond supergravity\mathoff}%
\bigskip\bigskip

\addtocounter{footnote}{1} 
\textbf{
N.~Beisert\footnote{\texttt{nbeisert@aei.mpg.de}}, 
M.~Bianchi\footnote{\texttt{Massimo.Bianchi@roma2.infn.it}}, 
J.~F.~Morales\footnote{\texttt{Francisco.Morales@lnf.infn.it}}, and
H.~Samtleben\footnote{\texttt{H.Samtleben@phys.uu.nl}}}

\addtocounter{footnote}{-3} \vspace{.3cm} $^\fnsymbol{footnote}$ 
\textit{Max-Planck-Institut f\"ur Gravitationsphysik\\
Albert-Einstein-Institut\\
Am M\"uhlenberg 1, D-14476 Potsdam, Germany}

\addtocounter{footnote}{1} \vspace{.3cm} $^\fnsymbol{footnote}$ 
\textit{
Dipartimento di Fisica and INFN \\
Universit\`a di Roma ``Tor Vergata''\\
00133 Rome, Italy}

\addtocounter{footnote}{1} \vspace{.3cm} $^\fnsymbol{footnote}$  
\textit{
Laboratori Nazionali di Frascati,\\
Via E. Fermi, 40\\ I-00044 Frascati (Rome), Italy}

\addtocounter{footnote}{1} \vspace{.3cm} $^\fnsymbol{footnote}$
\textit{Institute for Theoretical Physics and Spinoza Institute\\
Utrecht University, Postbus 80.195\\
3508 TD Utrecht, The Netherlands}

\setcounter{footnote}{0}

\end{center}
\bigskip

\begin{abstract}
We test the spectrum of string theory on $AdS_5\times S^5$ derived in
\arxivno{hep-th/0305052} against that of single-trace gauge invariant
operators in free ${\cal N}=4$ super Yang-Mills theory.  Masses of
string excitations at {\it critical tension} are derived by
extrapolating plane-wave frequencies at
$g_{\scriptscriptstyle\mathrm{YM}}=0$ down to finite $J$.  On the SYM
side, we present a systematic description of the spectrum of
single-trace operators and its reduction to $PSU(2,2|4)$
superconformal primaries via a refined Eratostenes' supersieve.  We
perform the comparison of the resulting SYM/string spectra of charges
and multiplicities order by order in the conformal dimension $\Delta$
up to $\Delta=10$ and find perfect agreement.  Interestingly, the
SYM/string massive spectrum exhibits a hidden symmetry structure
larger than expected, with bosonic subgroup $SO(10,2)$ and thirty-two
supercharges.
\end{abstract}


\setcounter{page}{0}

\newpage

\section{Introduction}
\label{sec:intro}

The strong form of Maldacena's conjecture \cite{Maldacena:1998re}
relates perturbative super Yang-Mills theory (SYM) to a higher spin
(HS) gravity theory in its broken phase. At strictly zero gauge
coupling $\gym=0$, the HS symmetry is recovered. According to
holography this suggests the existence of a ``critical $AdS$ radius''
where infinitely many gauge particles come down to zero mass and the
string spectrum on the highly curved $AdS_5\times S^5$ should coincide
with that of free $SU(N)$ ${\cal N} =4$ SYM theory. The dynamics at this point
is practically frozen if not for the `mixing' of single- with
multi-particle states which is suppressed by $g\indup{s}\approx 1/N$.
In \cite{Bianchi:2003wx} the spectrum of Kaluza-Klein (KK) descendants
of string excitations in $AdS_5\times S^5$ was derived. The results
were written in the deceivingly simple and suggestive form:
%
\begin{eqnarray}
{\cal H}_{\rm AdS} &=& {\cal H}_{\rm sugra}+\tkk
\, \tlong\, \sum_{\ell=1}^\infty\, 
({\rm vac}_\ell\times {\rm vac}_\ell)
\;,\label{flat}
\end{eqnarray}
where ${\rm vac}_\ell$ ($\ell=N_L=N_R$ denoting the string level)
encodes the physical spectrum of chiral string primaries in flat
space, properly rearranged in representations of $SO(10)$ as we shall
describe in this paper.\footnote{ The products in (\ref{flat}) are
understood in terms of the $SO(6)\times SO(4)$ subgroup.}  The
supergravity states ${\cal H}_{\rm sugra}$ organize into
$\frac{1}{2}$-BPS multiplets, while massive string excitations
($\ell\ge1$) sit in long multiplets given by tensoring string
primaries with the long Konishi supermultiplet $\tlong$
\cite{Andrianopoli:1998ut,Bianchi:2001cm} on $AdS$. Finally,
$\tkk=\sum_n (n00;00)$ is a polynomial of $SO(6)\times SO(4)$
representations accounting for Kaluza-Klein (KK) descendants.

\nocollect{Metsaev:1998it}%
\nocollect{Pesando:1998fv}%

Despite some progress
\cite{Berkovits:2002qx,Berkovits:2002zv,Berkovits:2000yr,
Berkovits:2000wm,Berkovits:1999xv,Berkovits:1999im,Metsaev:1998it,
Metsaev:2000yu,Metsaev:2000yf,Metsaev:2000bj,Pesando:1998fv,
Kallosh:1998nx}, a viable quantization scheme for type IIB string
theory on $AdS_5\times S^5$ is still lacking. Linearized field
equations around $AdS$ are not available and therefore assignments of
conformal dimensions $\Delta$ in (\ref{flat}) were missing.  In
\cite{Bianchi:2003wx}, we exploited the higher spin symmetry
enhancement in the {\it critical tension} limit\footnote{The study of
propagating strings on AdS near the HS critical radius has been
recently addressed in
\cite{Dhar:2003fi,Lindstrom:2003mg,Gopakumar:2003ns,Leigh:2003ez,Bonelli:2003zu,deMedeiros:2003hr}.} 
($R^2 \approx\alpha^\prime$) in order to fix the masses, \ie dimensions, of states
in the ``first Regge trajectory''\footnote{The quotation marks
distinguish this from the more familiar Regge trajectory
$\Delta=2+s_4$ with $s_4$ the four- (rather than the ten-)dimensional
spin.} $\Delta=2+s_{10}=2 \ell+n$ with string level $\ell$ and KK
floor $n$. In particular, massless HS gauge particles correspond to
states in the first Regge trajectory with $s_4 = 2\ell - 2$, and
$n=0$~\cite{Sezgin:2001zs,Sezgin:2001yf,Sezgin:2002rt}.  Scalar fields
dual to $\half$-BPS operators and their superpartners are associated
to KK recurrences of supergravity ($\ell =0, n\ge 2$) states.
Similarly, string primaries associated to marginal/relevant
deformations of ${\cal N}=4$ SYM were also found in the ``first Regge
trajectory'' ($\ell\le 2, n\le 4$) and the right spectrum of charges
and multiplicities was reproduced \cite{Bianchi:2003wx}.

The purpose of this paper is to refine the above analysis and present
a systematic description of the string/operator map at $\gym=0$ beyond
the first Regge trajectory. We will assign conformal dimensions of
states in (\ref{flat}) by setting first $\gym=0$ and then flowing to
the plane-wave limit~\cite{Berenstein:2002jq}. More precisely,
exploiting the manifest $SO(10)$ form of $\sum ({\rm vac}_\ell\times
{\rm vac}_\ell)$ and choosing an $SO(2)_J\times SO(8)$ subgroup
inside, we assign conformal dimensions $\Delta$ in such a way that:
\begin{eqnarray}
\Delta \,-\,J&=&\nu \;, \label{magic0}
\end{eqnarray}
with $\nu= \sum N_n w_n ~\to~ \sum_n N_n$ the plane-wave string
frequencies at $\gym=0$. This is to be contrasted with the opposite
limit, string theory in flat spacetime, where the Hamiltonian measures
the string level $\ell=\sum_n\,n N_n$. We will extrapolate formula
(\ref{magic0}) down to finite~$J$ along the line $\gym=0$. The perfect
match between the string spectrum of conformal dimensions found in
this way and that of SYM strongly suggests that this simple formula is
indeed exact at $\gym=0$ and we apply it to the full string spectrum.
This is a realization of the idea proposed in \cite{Beisert:2002tn,Metsaev:2002re}
that the BMN spectrum can be extended to the generic finite~$J$ case
and thus give the full spectrum a `stringy' nature.  The possibility
of assembling states in irreducible `massless' and `massive' HS
multiplets will only be hinted at in the conclusions and discussed in
greater details in a companion paper \cite{Beisert:2003xz}.

On the SYM side, the counting of gauge invariant single-trace
composite operators is performed in the framework of Polya
theory~\cite{Polya} that allows to count `words' of arbitrary length,
\ie `letters' in a given alphabet, modulo the action of the cyclic
group, for $SU(N)$, in order to account for the cyclicity of the trace.  This is
tantamount to the more romantic problem of counting `necklaces' made
of an arbitrary number of `beads' of given `colors'.

In order to reduce `entropy' in the comparison, we find it convenient
to focus on `superprimaries' of $PSU(2,2|4)$.  We achieve this goal by
refining the Eratostenes' Supersieve procedure introduced in
\cite{Bianchi:2003wx} and further getting rid of KK recurrences so as
to expose a hidden $SO(10,2)$ structure in the SYM spectrum beyond the
$\half$-BPS series $\zbps$: The AdS/CFT correspondence
predicts that the spectrum of gauge invariant ${\cal N}=4$ SYM
operators should take a form analogous to that of~\Ref{flat}
%
\begin{eqnarray}
  \zsym&=&
  \zbps  +  \mathcal{T}\indup{KK}\;  \tlong \;
  \zten \;.
\label{predictQ}
\end{eqnarray}
In particular, this implies that SYM states beyond the $\half$-BPS
series will be organized in towers of superconformal and
``Kaluza-Klein'' descendants. Remarkably, these two towers can be
combined into a single ``multiplet'' $\tten\equiv \tkk\; \tlong $,
which is generated by $SO(10)$ supertranslations $Q_{16},P_{10}$ with
$P_{10}^2=0$ (arising from the tracelessness condition for KK
harmonics), suggesting a hidden $SO(10,2)$ structure, cf.~\Ref{so10}
below.  Indeed, acting with $P_{10}$ lifts $SO(10)\times SO(2)$
representations to $SO(10,2)$. This points towards the extension of
the hidden 12-dimensional structures found in the KK towers of
$AdS_5\times S^5$ supergravity~\cite{Bars:2002pe} to the entire
massive string spectrum!  On the SYM side this implies that all SYM
states beyond the $\half$-BPS series can be organized in
representations of $SO(10,2)_{\Delta}$ rather than $SO(6)\times
SO(4,2)_{\Delta}$.\footnote{Which supergroup realizes this
``underlying symmetry'' is not clear to us.} We have explicitly
verified this structure till $\Delta=10$.

Even stronger, the AdS/CFT correspondence yields a quantitative prediction
obtained by comparing the spectra on the string \Ref{flat} and SYM
\Ref{predictQ} side:
\begin{eqnarray}
 \zten &\stackrel{!}{\equiv}& 
 \sum_{\ell=1}^\infty ({\rm vac}_\ell\times {\rm vac}_\ell)
\;.
\label{predict}
\end{eqnarray}
As the main result of this paper, we have verified this relation as
well up to $\Delta=10$, supporting the conjectured mass formula
\Ref{magic0}.

At small but finite 't~Hooft coupling, all but a handful of massless
states become massive in a Pantagruelic Higgs mechanism, whereby
higher spin multiplets in the bulk eat lower spin multiplets and
become `long' and `massive'. The counterpart of this `grande bouffe'
in the boundary theory is the appearance of anomalous dimensions
\cite{Bianchi:1999ge,Bianchi:2000hn,Dolan:2001tt,Bianchi:2001cm,
Bianchi:2002rw,Arutyunov:2002rs,Beisert:2002bb,Constable:2002vq} that,
following \cite{Beisert:2003jj}, can be systematically and almost
straightforwardly computed at one-loop for all gauge invariant
operators thanks to the integrability of the associated 
`super-spin chain' \cite{Minahan:2002ve,Beisert:2003yb,Dolan:2003uh}. This is
however beyond the scope of our analysis and will be taken care of (in
a restricted yet interesting set of states) in \cite{Beisert:2003xz}.
The related problem of establishing integrability beyond one-loop
\cite{Beisert:2003tq} is the subject of active investigation
\cite{Beisert:2003ea,Arutyunov:2003rg,Klose:2003qc,Beisert:2003ys} we
have little to say about here.

The plan of the paper is as follows. In section~\ref{sec:string}, we
review the naive procedure to compute the string spectrum on
$AdS_5\times S^5$ starting from the light-cone GS formalism in flat
space. Exploiting the hidden $SO(10)$ covariance and the BMN limit, we
propose a very simple yet effective mass/dimension formula for all string
states at the HS enhancement point dual to free SYM theory.  In
section~\ref{sec:N4}, we compute the SYM spectrum by resorting to
Polya theory, and describe how to identify HWS, \ie superprimaries, by
means of a refined Eratostenes' Supersieve. Removal of `KK
recurrences' further simplifies the problem of comparing the two
spectra.  We find perfect agreement for all the states we have
explicitly analyzed, \ie superprimaries with $\Delta\leq 10$ and their
descendants.  Section~\ref{sec:concl}, contains our conclusions and
perspectives. Various appendices tabulate our results and contain
unwieldy but necessary formulae for the interested reader.

\section{String states on $AdS_5\times S^5$}
\label{sec:string}

In this section, we review the KK reduction of string theory
on $AdS_5\times S^5$ following~\cite{Bianchi:2003wx}, and
propose a mass formula~\eqref{magic} for 
the entire string spectrum at {critical tension}, \ie at 
the HS enhancement point dual to $\gym=0$.  
To linear order in fluctuations around
$AdS_5\times S^5$ the type IIB field equations boil down to a set of
uncoupled free massive equations 
 \begin{eqnarray} 
 \left(\nabla_{AdS_5\times S^5}^2-M_{\Phi}^2 \right)
\Phi_{\cal R}&=&0 
\;,
\label{emh} 
\end{eqnarray} 
with ${\cal R}$ labeling irreducible representations of 
$SO(4,1)\times SO(5)$, the Lorentz group on $AdS_5\times S^5$, and running over the
spectrum of type IIB string excitations in flat space
\cite{Bianchi:2003wx}.  The form of \eqref{emh} is fixed by Lorentz
covariance, while ``masses'' $M_\Phi^2$ describe the coupling of
fields $\Phi_{{\cal R}}$ to the curvature and five-form flux. They can
in principle be determined by explicit evaluation of the linearized
equations around $AdS_5\times S^5$ but these equations are not
available beyond the supergravity level
\cite{Sezgin:2001zs,Sezgin:2001yf,Sezgin:2002rt}. Even if this
information is missing we will see how most of the information about
the string spectrum can be derived from standard KK techniques while
masses at {\it critical tension} can be fixed by requiring a
consistent plane-wave limit.

The spectrum of KK harmonics is determined by group
theory~\cite{Salam:1982xd} (see
\cite{deBoer:1998ip,deBoer:1998us,Gava:2001ne,
Gava:2002dx,Morales:2002ys,Morales:2002uh} for applications in the
$AdS$ context).  Expanding the ten-dimensional fields $\Phi_{{\cal
R}}$ in $S^5$-spherical harmonics ${\cal Y}^r_{{\cal R}_{SO(5)}}(y)$,
leads to
\begin{eqnarray}
\Phi_{{\cal R}}(x,y)&=&\sum_{ r\,\in\, {\rm KK}[{{\cal R}_{SO(5)}}] } 
{\cal X}^{ r}_{{\cal R}_{SO(1,4)}}(x)\, 
{\cal Y}^r_{{\cal R}_{SO(5)}}(y) 
\;,
\label{harm} 
\end{eqnarray}
with ${\cal R}={\cal R}_{SO(1,4)}\times {\cal R}_{SO(5)}$ and $x$, $y$ being 
coordinates along $AdS_5$ and $S^5$ respectively. The
sum runs over the set ${\rm KK}[{{\cal R}_{SO(5)}}]$ of all $SO(6)$
representations that contain ${\cal R}_{SO(5)}$ in their
decomposition under $SO(5)$.

This program was carried out in \cite{Bianchi:2003wx}.
The result for the massive string spectrum obtained this way
may be written as
\begin{eqnarray}
{\cal H}_{\rm AdS} &=& {\cal H}_{\rm sugra}+\tkk
\, \tlong\, \sum_{\ell=1}^\infty ({\rm vac}_\ell\times {\rm vac}_\ell)
\;, \label{kk}
\end{eqnarray}
with $\tkk$, $\tlong$ the KK and AdS superconformal descendant
 polynomials\footnote{The $SO(4,2)$ content $D(\Delta|j_1,j_2)$ can be
 read from that of $SO(4)\times SO(2)$ presented here by omitting
 $P_4$ translations. This was the notation adopted in
 \cite{Bianchi:2003wx}.}  :
 \begin{eqnarray}
\tkk&=& \sum_{n=0}^\infty  \; (n00;00)^n
\;,
\nn\\
\tlong &=& (1+Q+Q\wedge
Q+\ldots)\,(1+P_4+(P_4\times P_4)_s+\ldots ) \;,
 \nn\\
Q &=& (\half\,\half\,\half\,;\half\,\half)^{{\frac12}}\;,
\quad\quad P_4 ~=~ (000;10)^{1}\; ,
 \label{susys}
\end{eqnarray}
of $SO(6)\times SO(4)$ representations. Here, $(n00;00)$ refers to the
symmetric traceless representations of $SO(6)$, singlet of $SO(4)$, see
Appendix~\ref{app:notation} for our notation. AdS supermultiplets
$\tlong$ are instead generated by $16$ supersymmetries $Q$ and $4d$
derivatives $P_4$ along the $AdS_5$ boundary. Finally
$\sum_{\ell=1}^\infty ({\rm vac}_\ell\times {\rm vac}_\ell)$ denotes
the massive string spectrum in flat space, rearranged in terms of
$SO(6)\times SO(4)$ representations --- this is achieved by lifting
the original $SO(9)$ representations up to $SO(10)$, breaking down to
$SO(6)\times SO(4)$ --- and supplied it with a mass quantum number
$\Delta$, denoted by a superscript ${\cal R}^\Delta$.

The spectrum \eqref{kk} thus naturally assembles into long
multiplets of $PSU(2,2|4)$
\begin{eqnarray}
{\cal A}^\Delta_{(w_1,w_2,w_3;w_4,w_5)} &\equiv&
(w_1,w_2,w_3;w_4,w_5)^\Delta \times \tlong\;.
\label{multA}
\end{eqnarray}
At this point, the assignments of conformal dimensions are {\it ad
hoc};  they will be justified later on in this section.

\subsection{Strings on flat space, yet again}

For completeness, we review here the spectrum of type II superstrings
in flat space in the GS formulation.
Chiral (say left-moving) string excitations are created by the
raising modes  $\alpha_{-n}^I, S^{a}_{-n}$ acting on the
vacuum $|{\cal Q}_c\rangle$:
\begin{eqnarray}
 S^{a}_{n}|{\cal Q}_c\rangle &=&
\alpha^I_n|{\cal Q}_c\rangle~=~0 \;, \quad\quad\quad n>0 \;.
\end{eqnarray}
Here and below, indices $I=1, \dots, {\bf 8_v}$, $a=1, \dots, {\bf 8_s}$, 
$\dot{a}=1, \dots, {\bf 8_c}$ run over the vector, spinor left
and spinor right representations of the $SO(8)$ little Lorentz
group. In addition we have introduced the compact notations:
\[ {\cal Q}_s~=~{\bf 8_v}-{\bf 8_s}\;,
\quad\quad {\cal Q}_c~ =~{\bf 8_v}-{\bf 8_c}
\;,
\label{8vs}
\]
to describe chiral worldsheet supermultiplets. The vacuum $|{\cal
Q}_c\rangle$ is $2^4$-fold degenerate as a result of the
quantization of the eight fermionic zero modes $S^{a}_{0}$. The
physical spectrum of the type IIB  superstring is defined by
tensoring two (left and right moving) identical chiral spectra
subject to the level matching condition $\ell\equiv N_L=N_R$.  
The spectrum $T_\ell$ at
 string excitation level $\ell$ takes the form
\begin{eqnarray}
T_\ell &=& {\cal Q}_c^2 \;\times \,
\Big(\sum_{\sum r \,k_r =\ell} \prod_r {\cal Q}_s^{{\bf\cdot}
k_r}
\Big)^2 \;,
\label{TQ}
\end{eqnarray}
where we use the notation of  dotted
(graded symmetrized) products according to
\begin{eqnarray}
{\cal Q}_s^{\cdot 2}
&\equiv & {\bf 8}_{(v}\times 
{\bf 8}_{v)}+{\bf 8}_{[s}\times {\bf 8}_{s]}-{\bf 8}_v\times
{\bf 8}_s~=~{\bf 8}_v\times {\cal Q}_s \;,\nn\\
{\cal Q}_s^{\cdot3} &\equiv & {\bf 8}_{(v}\times {\bf 8}_v\times
{\bf 8}_{v)}-{\bf 8}_{[s}\times {\bf 8}_s\times  
{\bf 8}_{s]}+{\bf 8}_v\times {\bf 8}_{[s}\times
{\bf 8}_{s]}-{\bf 8}_s\times {\bf 8}_{(v}\times {\bf 8}_{v)} \nn\\
&=& ({\bf 35}-{\bf 8_c})\times {\cal Q}_s \;.\non
\mbox{etc.}
\label{fcs}
\end{eqnarray}
(In contrast, `$\times$' and `$\prod$' refer to the ordinary
tensor product.) For the dotted products we further find the
following recursive formula
\begin{eqnarray}
{\cal Q}_s^{\cdot n} &=& \Xi_n \times {\cal Q}_s \;,
\label{Qrec}
\end{eqnarray}
with
$$
\Xi_n ~=~ {\cal L}(\Xi_{n-1}) +
\left\{
\arraycolsep3pt
\begin{array}{ll}
-{\bf 8_c} &n=3\\
{\bf 8_v} &n=4\\
-{\bf 8_c}+{\bf 28} &n=5\\
{\bf 8_v}-{\bf 8_s} &n>5\;\; \mbox{even}\\
{\bf 1}-{\bf 8_c}+{\bf 28} &n>5\;\; \mbox{odd}\\
\end{array}\right.
\;,
\;\;
{\cal L}([k,l,p,q])\equiv [k\pls1,l,p,q]\;.
$$
For the first massive levels we then find explicitly
\begin{eqnarray}
T_1 &=& {\cal Q}_c^2 \times {\cal Q}_s^2=
T_1\times ({\bf 1}_1)^2  \;,\label{T123}\\[.5ex]
T_2 &=& {\cal Q}_c^2\times ({\cal Q}_s+{\cal Q}_s\cdot {\cal Q}_s )^2 =
T_1\times ({\bf 1}_1+{\bf 8_v}_{,2})^2  \;, \nn\\[.5ex]
 T_3 &=& {\cal Q}_c^2\times ({\cal Q}_s+{\cal Q}_s^2+{\cal Q}_s\cdot {\cal
Q}_s\cdot   {\cal Q}_s)^2=
  T_1 \times ({\bf 1}_1+{\bf 8_v}_{,2}-{\bf 8_s}_{,2}+{\bf 35_v}_{,3}-
  {\bf 8_c}_{,3})^2
\;;\nn
\end{eqnarray}
for general $\ell>0$, the structure is
\begin{eqnarray}
T_\ell &\equiv&
 T_1 \times ({\rm vac}_\ell\times {\rm vac}_\ell) ~=~
 T_1 \times ({\textstyle \sum}_j {\cal R}_{j,\nu})^2
\label{flatvac}
\;,
\end{eqnarray}
where the $SO(8)$ content of the chiral ground states ${\rm vac}_\ell$
is found by explicitly evaluating \eqref{TQ}, \eqref{Qrec}. We have
introduced the subscript ``$\nu$'' for each $SO(8)$ representation
${\cal R}_j$, indicating the excitation number, \ie the number of
${\cal Q}_s=\{ \alpha^I_{-n}, S^a_{-n} \}$ that generate the state
${\cal R}_{j,\nu}$ from the vacuum. Notice that disregarding this
quantum number, the massive spectrum naturally assembles into $SO(9)$
representations, as expected.

To proceed with the string spectrum on $AdS_5\times S^5$, we note that
the massive string spectrum in flat space may in fact be further
lifted from $SO(9)$ to $SO(10)$ in terms of representations
\begin{eqnarray}
[k,l,m,p,q]^* &\equiv& [k,l,m,p,q]-[k\!-\!1,l,m,p,q]
\;,\qquad (k>0) \;.
\end{eqnarray}
For the first few string levels described by (\ref{T123}) one finds
\footnote{ Bosonic (fermionic) states with negative (positive)
multiplicities in ${\rm vac}_\ell$ represent unphysical degrees of
freedom.  This minus sign, contrary to minus signs in previous
formulae, is not related to spin and statistics.  Note that after
multiplying $({\rm vac}_\ell\times {\rm vac}_\ell)$ with the KK tower
according to \eqref{kk}, the unphysical states drop
out~\cite{Bianchi:2003wx}.}
\begin{eqnarray}
{\rm vac}_1 &=& [0,0,0,0,0]^1 \;,
\nn\\
{\rm vac}_2 &=& [1,0,0,0,0]^2-[0,0,0,0,0]^3 \;,
\nn\\
{\rm vac}_3 &=& [2,0,0,0,0]^3-[1,0,0,0,0]^4+[0,0,0,0,1]^{5/2} \;,
\label{v123}
\end{eqnarray}
with $[k,l,m,p,q]$ the Dynkin labels of $SO(10)$, and the superscript
$\Delta_L$ referring to the chiral contribution to the conformal
dimensions.  Here and in the following, $\Delta=\Delta_L+\Delta_R$
will always refer to the bare dimension, \ie at $\gym=0$, where the
conformal dimensions (in general a complicated function of the $AdS$
radius) reduce to integer or half-integer numbers. At this stage there
is still an ambiguity in the lift to $SO(10)$ due to the different
spinor chiralities of $SO(10)$. We shall comment on this in the next
subsection.

The assignments for conformal dimensions $\Delta_L$ in \eqref{v123}
have been chosen {\em ad hoc} such that the physical states saturate
the $SO(10)$ bound
\begin{eqnarray}
\Delta_L [k,l,m,p,q] &\ge& 1+k+2 l+3 m+\sfrac{5}{2} p + \sfrac{3}{2} q
\;. 
\label{bound}
\end{eqnarray}
This bound emerges from the $PSU(2,2|4)$ unitarity
bound~\cite{Dobrev:1985qv} after breaking $SO(10)$ down to
$SO(6)\times SO(4)\subset PSU(2,2|4)$.  The assignments \Ref{v123}
will be justified in the next subsection in a more general setting
where string primaries with conformal dimensions beyond the unitarity
bound will also be found at higher string levels.

Let us close this section by observing a peculiarity of the massive
flat space string spectrum: the partial sums of chiral string
excitations $\sum_{\ell=1}^{\ell_m} \; {\rm vac}_\ell $ naturally
assemble into true $SO(10)$ representations! E.g.  ${\rm vac}_1 = {\bf
1}$, ${\rm vac}_1+{\rm vac}_2 = {\bf 10}$, ${\rm vac}_1+{\rm
vac}_2+{\rm vac}_3 = {\bf 54} -{\bf 16_s}$, and so on.

\subsection{Mass formula}

According to \eqref{kk}, the string spectrum on $AdS_5\times S^5$ may
be derived from that of string theory in flat space, upon multiplying
in the KK towers.  To this end, the $SO(10)$ content~\eqref{v123} of
the chiral ground states ${\rm vac}_\ell$ is supplied with an additional $SO(2)$ quantum
number $\Delta$, the conformal dimension. Upon breaking $SO(10)$ down
to $SO(6)\times SO(4)$, they yield the highest weight states of the
long multiplets \Ref{multA} in which the string spectrum is
organized. In this subsection, we will specify the conformal
dimensions associated with the $SO(10)$ content of string
primaries. We focus on the HS symmetry enhancement (free SYM) point.
  
As discussed in the introduction, the free SYM limit is accessible
from the plane-wave regime, where
a mass formula describing conformal dimensions of operators
with a large $SO(2)_J\in SO(6)$ charge $J$ is available
 \cite{Berenstein:2002jq,Gross:2002su,Santambrogio:2002sb}
\begin{eqnarray}
\Delta -J &=&  \sum_n N_n w_n ~\equiv~ \sum_n N_n
\sqrt{1+\frac{\gym^2 N n^2}{J^2}}
\;,
\label{bmn}
\end{eqnarray}
Following the conjecture of a stringy spectrum even 
outside the BMN regime \cite{Beisert:2002tn,Metsaev:2002re}
we will argue that at $\gym=0$, the masses for the entire string spectrum 
can be fixed by extrapolating this formula down to finite $J$. 
 
Specifically, breaking the $SO(10)$ spectrum \eqref{v123} down to
$SO(8)\times SO(2)_J$ yields states $[k,l,p,q]^\Delta_J$. These
are to be identified with the string state $[k,l,p,q]_{\nu}$ with
$\nu$ the string excitation number introduced in~\eqref{T123} above.
At $\gym=0$ the BMN mass formula \eqref{bmn} then reads
\footnote{This
limit is to be contrasted with that of string
theory in flat space, where
the string Hamiltonian measures the occupation number $\ell=\sum_n n
N_n$ rather than $\nu=\sum_n N_n$. Note that the level matching
condition $\ell=\bar{\ell}$ in contrast is left unaltered by sending $\gym$
to zero. }
\[
\Delta -J~=~ \nu ~\equiv~ \sum_n N_n\;.
\label{magic}
\]
We hence propose this relation to hold on the entire massive
string spectrum at $\gym=0$. Given the $SO(10)$ content of
the flat space string spectrum~\eqref{flatvac}, equation~\eqref{magic}
uniquely determines the bare dimensions $\Delta$ in the string
spectrum on $AdS_5\times S^5$. As an illustration, let us consider
the first few string levels given above:
\<
{\rm vac}_1 \eq [0,0,0,0,0]^1 \non
\earel{\stackrel{SO(8)\times SO(2)}{\rightarrow}} [0,0,0,0]^1_0
\non
\earel{\stackrel{\eqref{magic}}{\rightarrow}} [0,0,0,0]_1
\;,
\nn\\[1.5ex]
{\rm vac}_2 \eq [1,0,0,0,0]^2-[0,0,0,0,0]^3 \non
\earel{\stackrel{SO(8)\times SO(2)}{\rightarrow}}
 [1,0,0,0]^2_0+
[0,0,0,0]^2_1+[0,0,0,0]^2_{-1} -[0,0,0,0]^3_0\label{second}\\
\earel{\stackrel{\eqref{magic}}{\rightarrow}}
 [1,0,0,0]_2+ [0,0,0,0]_1 \;,
 \nn\\[1.5ex]
{\rm vac}_3 \eq [2,0,0,0,0]^3-[1,0,0,0,0]^4-[0,0,0,0,1]^{5/2} \non
\earel{\stackrel{SO(8)\times SO(2)}{\rightarrow}}
  [2,0,0,0]^3_0+
[1,0,0,0]^3_1+[1,0,0,0]^3_{-1}
+[0,0,0,0]^3_0+ [0,0,0,0]^3_2 \nl +[0,0,0,0]^3_{-2}
 -  [1,0,0,0]^4_0- [0,0,0,0]^4_1-[0,0,0,0]^4_{-1} \nl
-[0,0,0,1]^{5/2}_{1/2}-[0,0,1,0]^{5/2}_{-1/2} \non
\earel{\stackrel{\eqref{magic}}{\rightarrow}}
 [2,0,0,0]_3 +[1,0,0,0]_2
+[0,0,0,0]_1
-[0,0,1,0]_{3}-[0,0,0,1]_{2}
\;,
\nn
\>
which precisely reproduces \eqref{T123} and thus confirms our
{\em ad hoc} assignments \eqref{v123}! For these low massive
levels, the conformal dimensions determined by~\eqref{magic}
all saturate the unitarity bound~\eqref{bound}. At higher
levels, starting from the $\Delta=3$ singlet at level
$\ell=5$, this bound is still satisfied but no longer
saturated; the correct conformal dimensions are rather
obtained from \eqref{magic}.

To summarize, the massive flat space string spectrum may be lifted to
$SO(10)\times SO(2)_\Delta$, such that breaking $SO(10)$ down to
$SO(8)\times SO(2)_J$ reproduces the original $SO(8)$ string spectrum
and its excitation numbers \eqref{T123} via the relation
\eqref{magic}. The results up to string level $\ell=10$ are displayed
in Appendix~\ref{app:string}. For these levels, relation~\eqref{magic}
not only determines the conformal dimensions $\Delta$, but also fixes
all ambiguities, arising in the lift of the $SO(9)$ massive string
spectrum to $SO(10)$. Here, superconformal and KK descendants are
included by replacing the flat long multiplet $T_1$ by the product of
the long Konishi multiplet $\tlong$ and the tower of KK descendants
$\tkk$
\begin{equation}
T_1 ~\to~ \tkk\,\tlong \;.
\end{equation}
Eventually, breaking this $SO(10)$ down to $SO(6)\times SO(4)$ then
yields the massive string spectrum on $AdS_5\times S^5$ via
\eqref{kk}. Based on the relation~\eqref{magic}, we have thus been
able to assign the string masses at $\gym=0$. It would clearly be
interesting, to achieve a direct derivation of this result.  In
Appendix~\ref{app:stringpart}, we have formulated some of these
results in terms of the string partition function. In the next section
we shall compare the string spectrum to ${\cal N}=4$ SYM theory.

 \section{${\cal N}=4$ SYM theory}
\label{sec:N4}

The spectrum of KK descendants of fundamental string states can be
precisely tested against that of gauge invariant operators in $SU(N)$ ${\cal
N}=4$ SYM theory.  In this subsection we apply Polya theory
\cite{Polya,Sundborg:1999ue,Polyakov:2001af}, (see
\cite{Bianchi:2003wx} for a quick review) and the Super-sieve
algorithm introduced in \cite{Bianchi:2003wx} to determine the
spectrum of superconformal primaries in ${\cal N}=4$ SYM theory.  To
facilitate the reading of this section, we first explain the algorithm
in full generality and display in section \ref{sec:3.3} explicit
expressions only for the `blind partition function', that counts
states according to their conformal dimensions, independently of the
remaining quantum numbers.  The full spectrum of multiplicities and
quantum numbers will be displayed in the appendix.

 \subsection{The single trace spectrum}

The spectrum of gauge invariant ${\cal N}=4$ SYM operators is
organized under the supergroup $PSU(2,2|4)$.  Multiplicities will be
encoded in the weighted partition function:
  \begin{eqnarray}
\mathcal{Z}_n(t,y_i)={\rm Tr}_{n\mathrm{-letters}} (-)^F\,
t^{\Delta}\, 
{\bf y}^{\bf w}\;,
 \quad\quad {\bf y}^{\bf w}\equiv \prod_{i=1}^5 y_i^{w_i}
\;,
 \label{zy}
 \end{eqnarray}
where the vector ${\bf w}=(w_1,w_2,w_3;w_4,w_5)$ and $\Delta$
label the charges under the canonical $SO(2)^3\times
SO(2)^2\times SO(2)_\Delta$ Cartan generators of the
$SO(6)\times SO(4)\times SO(2)_\Delta$
compact bosonic subgroup of $PSU(2,2|4)$. The insertion $(-)^F$,
$F$ being the fermion number, accounts for the right spin
statistics. We restrict ourselves to single trace operators,
\ie\, for $SU(N)$, cyclic words built from SYM letters: $\phi^i$,
$\lambda^A_\alpha$, $\lambda_{A \dot{\alpha}}$, $F_{\mu\nu}$ and
derivatives 
thereof.\footnote{The indices $i=1,\dots, 6$, 
$\mu=0,\dots, 3$, label the vector representations of $SO(6)$ 
and $SO(4)$, while $A=1,\dots, 4$, and $\alpha,\dot{\alpha}=1,2$
label the spinor representations.}

A word consisting of a single letter is clearly cyclic. The
field content of single-letter words is then encoded in
\<\label{sing1}
\mathcal{Z}_1(t,y_i) \eq \sum_{s=0}^\infty \, \left[
t^{s+1} \,\partial^s \phi+t^{s+\frac{3}{ 2}}\,\partial^s \lambda +
t^{s+\frac{3}{ 2}}\,\partial^s \bar\lambda +
t^{s+2}\,\partial^s F  \right]
\nln
\eq \sum_{s=0}^\infty \, \left[
\chi_{(100)}\chi_{(s,0)}\, t^{s+1}+
\chi_{(000)}\chi_{(s+1,1)}\, t^{s+2}
+\chi_{(000)}\chi_{(s+1,-1)}\, t^{s+2}
\right.
\nl
\qquad\qquad \left.
-\chi_{(\frac{1}{2},\frac{1}{2},
\frac{1}{2})}\chi_{(s+\frac{1}{2}, \frac{1}{2})}\,
t^{s+\frac{3}{ 2}}
-\chi_{(\frac{1}{2},\frac{1}{2},
-
\frac{1}{2})}\chi_{(s+\frac{1}{2},-\frac{1}{2})}\, t^{s+\frac{3}{ 2}}
\right]
\;,
\label{z1}
\>
where $\chi_{(w_1,w_2,w_3)}\chi_{(w_4,w_5)}$ denotes
the character polynomials of $SO(6)\times SO(4)$
representation with highest weight state
$(w_1,w_2,w_3;w_4,w_5)$. They are given by 
formulas (\ref{so6}),(\ref{so4}) in Appendix~\ref{app:notation}.
Cyclic words with $n>1$ letters are given in terms of Polya's 
formula~\cite{Bianchi:2003wx}
\<
\zsym(t,y_i)\eq\sum_{n=2}^\infty
\, \mathcal{Z}_n (t,y_i)= \sum_{n,n|d} 
\frac{\varphi(d)}{n}\,{\cal
Z}_1(t^d,y^d)^{\frac{n}{d}} \;,
\label{polyah}
\> with the sum running over all integers $n$ and their divisors $d$,
and Euler's totient function $\varphi(d)$ equals the number of
integers relatively prime to $d$ and smaller than $d$ with $\varphi(1)=1$ by
definition. The omission of the $n=1$ term in the sum is due to the
fact that we are considering SYM with gauge group $SU(N)$ rather than
$U(N)$.
 
The set of states in \eqref{polyah} organizes into multiplets of the
${\cal N}=4$ superconformal algebra $PSU(2,2|4)$. The spectrum of
superconformal primaries can be found by filtering \eqref{polyah} by a
sort of Eratostenes' Sieve, that removes at each step all descendants
from superconformal primaries and declare ``primaries'' the remaining
lowest conformal dimension states~\cite{Bianchi:2003wx}.
 
As the first step, we identify superconformal
primaries in $\zsym(t,y_i)$.
According to \eqref{multA}, the character polynomial
$\chi_{\wfat}^{\Delta}$ of a generic long supermultiplet 
of $PSU(2,2|4)$ with highest weight state
$t^{\Delta}\,{\bf y}^{\wfat} $ is generated by all
(super)translations in the following way (Racah-Speiser)
\[
 \chi_{\wfat}^{\Delta}
 (t,y_i)=\tlong(t,y_i)\, t^{\Delta}\,\chi_{\wfat}(y_i)\label{mlong}
 \;,
 \]
 where
\[
\tlong(t,y_i)=
\frac{\prod_{s=1}^{16}(1-t^\frac{1}{2} {\bf y}^{\wfat_{s}})}{
 \prod_{v=1}^{4}(1-t\, {\bf y}^{\wfat_{v}})} \;,
 \label{TT}
 \]
denotes the character polynomial of the long supermultiplet
$\tlong$ from \eqref{susys}. The vectors $\wfat_{s},\wfat_{v}$ in 
\eqref{TT} run over the weights of the 16-spinor and the 4-vector
representation of the supersymmetry Q and translation generator
P, respectively,
 \<
 \wfat_{s=1\ldots 16} \eq 
(\pm \half,\pm \half, \pm \half;\pm
\half,\pm \half)\;,\qquad \mbox{with }\prod_{i=1}^5 w_{s,i}>0\; ,
\nln
\wfat_{v=1\ldots 4} \eq (0,0,0;\pm 1,0),(0,0,0;0,\pm 1)\;.
 \label{weights}
\>

When interactions are turned on (but still at large $N$ to avoid
mixing with multi-trace operators for which further shortening
conditions may apply), single-trace operators fall into two classes of
$PSU(2,2|4)$ multiplets: $\half$-BPS and long multiplets
\cite{Bianchi:2003wx}:
 \[
  \zsym(t,y_i)= \zbps(t,y_i)+
    \zlong(t,y_i) \;,
 \label{zmult}
 \]
where
\[
\zbps(t,y_i)=\sum_{n=2}^\infty
\chi^n_{(n00;00)}(t,y_i) \;,
\label{ZBPS}
\]
is the partition function counting $\half$-BPS states and their
(super)descendants which corresponds to the supergravity spectrum.
The remaining $\zlong(t,y_i)$ is the main subject of our
investigations. States in $\zlong(t,y_i)$ sit in long multiplets of
the superconformal algebra $PSU(2,2|4)$.  This matches the multiplet
structure of the string result (\ref{kk}) and therefore comparisons to
string theory can be restricted to superconformal primaries.
Superconformal primaries can be found by factoring out $\tlong$ in
\eqref{zmult}:
\[\label{z46}
\zprim(t,y_i)=
\zlong(t,y_i) \,/\;
\tlong(t,y_i)
\;.
\]

\subsection{Comparison to the string spectrum}

The AdS/CFT correspondence predicts that the spectrum of
superconformal primaries in $SU(N)$ ${\cal N}=4$ SYM matches that of
string theory on $AdS_5\times S^5$:
\[
\zprim
~=~\tkk \;
\sum_{\ell=1}^\infty\,
({\rm vac}_\ell\times {\rm vac}_\ell)
\;,
\label{prediction}
\]
with
\<
\tkk \earel{\equiv}
\sum_{n=0}^\infty t^n\chi_{(n00)}=
(1-t^2)\,\prod_{v=1}^{6}(1-t\, {\bf y}^{{\bf q}_v})^{-1}
 \;,
\nl
{\bf q}_{v=1\ldots 6} =
(\pm1,0,0;00),(0,\pm1,0;00),(0,0,\pm1;00)
\;.
 \>
The
comparison is considerably simplified if one factorizes the
contribution of ``KK descendants'' from both sides in
\eqref{prediction}. We denote by $\zten$ the
partition function of SYM superconformal primaries up to KK
recurrences and $\half$-BPS states. According to \eqref{z46} one
finds:
\[ \label{zso10}
\zten
~\equiv~ \zprim\,/\;\tkk 
~\equiv~\zlong\,/\; \tten \;.
\]
 Remarkably, KK and superconformal descendants can be combined together into
a manifestly $SO(10)\times SO(2)$ covariant ``supermultiplet'':
\[ \tten=\tkk\tlong=(1-t^2)\,
\frac{\prod_{s=1}^{16}(1-
t^\frac{1}{2} \yfat^{\wfat_{s}})\, }{
 \prod_{v=1}^{10}(1-t\, \yfat^{\wfat_{v}})}
 \;.
\label{so10}
 \]
The appearance of $\tten$ that naturally extends the SCA to account
for KK descendants is rather suggestive.  Translations $P_{10}$ in
$\tten$ now carry the weights of a ten-dimensional vector in contrast
with its four dimensional cousins in \eqref{mlong}. The factor
$(1-t^2)$ subtracts the trace of KK recurrences and lifts to
$P_{10}^2=0$.  $\tten$ can be thought as the defining ``Konishi-like''
multiplet of a larger superalgebra with $SO(10,2)$ bosonic generators
and thirty-two supercharges.  Notice that since $\tten$ is manifestly
covariant under $SO(10,2)$, the full massive SYM spectrum
can be rearranged into representations of $SO(10,2)$ rather than
$SO(6)\times SO(4,2)$ if $\zten(t,y_i)$ does.  We have explicitly
checked that this is the case for all SYM primaries with $\Delta\leq
10$. The physical implications of this symmetry structure remain to be
explored.

The AdS/CFT prediction 
can now be stated in the simple form
\[
\zten\stackrel{!}{=}
\sum_{\ell=1}^\infty \, ({\rm vac}_\ell\times {\rm vac}_\ell) \;.
\label{PPrediction}
\]
This can be verified order by order in $\Delta$. In particular,
to order $\Delta_{\rm max}$ only terms in \eqref{polyah} with
$n<\Delta_{\rm max}$ letters contribute. We have carried out this
program till $\Delta_{\rm max}=10$. The spectrum of superconformal
primaries is displayed in Appendix~\ref{app:specN4} (however due to
the limited space only until $\Delta=\frac{13}{2}$).
Comparison with the string theory results for ${\rm
vac}_\ell$ with $\ell\leq 10$ collected in Appendix~\ref{app:string}
show perfect agreement till $\Delta_{\rm max}=10$~! These
results strongly support the conjectured dimension
formula~\eqref{magic}.

\subsection{SYM partition function}
\label{sec:3.3}

In this subsection, we illustrate our algorithm by focusing on the
`blind partition function' $\zsym(t)\equiv \zsym(t,1)$. The full
spectrum of SYM charges and multiplicities is obtained in a similar
way and the results are displayed in Appendix~\ref{app:specN4}.  The
starting point is the one-letter partition function (\ref{sing1}) at
$y_i=1$:
\<\label{sing1A}
\mathcal{Z}_1(t) \eq  \frac{2t(3+\sqrt{t})}{(1+\sqrt{t})^3}
\;.
\>
Plugging this in (\ref{polyah}) one finds
 \<
 \zsym(t)\eq\sum_{n=2}^\infty \sum_{n|d}
\frac{\varphi(d)}{n}\,\left[
\frac{2t(3+t^{\frac{d}{2}})}{(1+t^{\frac{d}{2}})^3}\right]^{\frac{n}{d}}
\label{polyahA}\\
\eq
21\,t^2 - 96\,t^{\frac{5}{2}} + 
376\,t^3 - 1344\,t^{\frac{7}{2}}
+ 4605\,t^4 - 15456\,t^{\frac{9}{2}} +
  52152\,t^5
\nl
- 177600\,t^{\frac{11}{2}}
+ 608365\,t^6 - 2095584\,t^{\frac{13}{2}} + 7262256\,t^7
- 25299744\,t^{\frac{15}{2}}
\nl
+ 88521741\,t^8 - 310927104\,t^{\frac{17}{2}} + 1095923200\,t^9
- 3874803840\,t^{\frac{19}{2}}
\nl
+ 13737944493\,t^{10}
+ \mathcal{O}(t^{\frac{21}{2}}) \;.
\>
The next step in our program is to identify superconformal primaries in
$\zsym(t)$. This can be easily done by first subtracting from
(\ref{polyahA}) states in the 1/2 BPS series~\eqref{ZBPS}
\<
\zbps(t)\eq
 \sum_{n=2} ^\infty
\frac{t^n\bigbrk{n+2-(n-2)t^{\frac{1}{2}}}}{12(1+t^{\frac{1}{2}})^4}
\Bigl[
(n+1)(n+3)\bigbrk{(n+2) -3t^{\frac{1}{2}}(n-2)}+ 
\nl\hspace{4.5cm}{}
+t(n-1)(n-3)\bigbrk{3(n+2)-t^{\frac{1}{2}}(n-2)}
\Bigr]
\nln
\eq
\frac{t^2\bigbrk{20+80t^{\frac{1}{2}}
+146t+144t^{\frac{3}{2}}+81t^2+24t^{\frac{5}{2}}+3t^3}}{(1-t)
(1+t^{\frac{1}{2}})^8}
\;,
\>
and then the KK and supersymmetry descendants, \ie dividing by
\[ \tten= (1-t^2)\frac{(1-t^\frac{1}{2})^{16}}{
 (1-t)^{10}}.
 \]
One finds
 \<
\zten(t)\eq \left[
  \zsym(t)-\zbps(t)\right]/
\tten(t)\label{numso10}\\
\eq
 t^2 + 100\,t^4 + 236\,t^5 - 1728\,t^{\frac{11}{2}} + 4943\,t^6
- 12928\,t^{\frac{13}{2}}
\nl
+ 60428\,t^7 -   201792\,t^{\frac{15}{2}}
 + 707426\,t^8 - 2550208\,t^{\frac{17}{2}}
\nl
+ 9101288\,t^9
- 32568832\,t^{\frac{19}{2}} + 116831861\,t^{10} +
\mathcal{O}(t^{\frac{21}{2}})
\;.\nn
\label{zso10A}
\>
 The expansion (\ref{numso10}) can be rewritten as
\<
\zten(t)\eq t^2+(10 t^2-t^3)^2+(-16 t^{5/2} + 54 t^3 - 10 t^4)^2\nn\\
&& +
(45 t^3 - 144 t^{\frac{7}{2}} + 210 t^4 + 16 t^{\frac{9}{2}} - 54 t^5)^2+\ldots
\;,
\>
etc., in perfect agreement with (\ref{PPrediction}) and the
string spectrum \eqref{v123}, Appendix~\ref{app:string}.

\section{Discussion}
\label{sec:concl}

We have analyzed the spectrum of string theory on $AdS_5\times S^5$ at
the HS enhancement point, exploiting the mass formula \Ref{magic}, and
compared to that of single-trace gauge invariant operators in free $SU(N)$
${\cal N}=4$ SYM theory. Up to $\Delta_{\rm max}=10$ we
found that the spectrum indeed organizes into $SO(10,2)$
representations, and confirms the AdS/CFT prediction
\Ref{predictQ}. There are additional SYM quantum numbers at $\gym =0$
whose string origin is not completely clear.

Classical type IIB
supergravity is invariant under the $U(1)_B$ compact `isotropy' subgroup of $SL(2,\mathbb{R})$ that acts by chiral transformations on the fermions. This symmetry
is anomalous at one-loop and is broken by string
interactions. However, as originally observed by Intriligator
\cite{Intriligator:1998ig,Intriligator:1999ff},
the structure of the type IIB effective
action suggests that amplitudes with up to four external supergravity
states should be $U(1)_B$ invariant even after including higher derivative terms
that receive non-perturbative corrections
\cite{Green:1997tv,Bianchi:1998nk,Green:2002vf,Kovacs:2003rt}.  In order
for this 'bonus symmetry' to be present at tree level, the exchange of massive string
states should not spoil it. This in turn implies that consistent
assignments of $U(1)_B$ charges to massive string states should be
feasible in principle, \eg by identifying allowed decay channels into
two supergravity states with known $U(1)_B$ charges. Three-point
amplitudes with all massive states would however violate $U(1)_B$.
From the holographically dual viewpoint of ${\cal N}=4$ SYM, $U(1)_B$
is an external automorphism of the superconformal algebra that extends
$PSU(2,2|4)$ to $SU(2,2|4)$ \cite{Heslop:2003xu}. It acts as a chiral transformation of the
four gaugini combined with a duality rotation of the field-strengths
and as such it cannot possibly be a symmetry of the theory at $\gym
\neq 0$ if not for a restricted class of correlation functions, \ie
any 2-point functions, 3-point functions with at most one operator in
a long multiplet and 4-point functions of operators in $\half$-BPS
multiplets.

The story of another quantum number, the length $L$ of a single-trace
operator is even more obscure. In SYM perturbation theory it is a
perfectly good quantum number up to and including one-loop.  The
(generalized) Konishi anomaly and other anomalous effects
\cite{Eden:2003sj,Beisert:2003ys,Bianchi:2003yy} imply the mixing of
operators with different lengths, \ie different number of
constituents, even at large $N$. Instanton effects, which are however
highly suppressed at large $N$, wash out any memory of this quantum
number not differently from larger orders in perturbation theory so it
is very hard to say how one could even in principle assign this
quantum number to string states before the HS enhancement point is
reached.\footnote{Nevertheless, $B$ and $L$ seem to be related in some
sense and possibly the combinations $L\pm B$ can be associated to the
left- and right-moving sectors of string theory, respectively.}

Other discrete quantum numbers allow for a more direct string
interpretation.  The parity $P$ identified in \cite{Beisert:2003tq} 
in the SYM spectrum is nothing but worldsheet parity $\Omega$ that 
survives as a symmetry
after compactification on $S^5$ and is `gauged' when $S^5$ is replaced
by $RP^5$ with or without discrete two-form fluxes
\cite{Witten:1998xy,Witten:1998bs,Bianchi:1992eu,Bianchi:1998rf}.  The
holographic counterpart of the orientifold projection is the breaking
of the $SU(N)$ gauge group to $SO(2N)$, ${SO(2N+1)}$ or $Sp(2N)$,
depending on the choice of two-form flux
\cite{Witten:1998xy,Witten:1998bs}.  We have confirmed the
correspondence of $P$ and $\Omega$ by including parity quantum numbers
in \eqref{PPrediction}.  In string theory, parity is obtained by the
formula
\[
({\rm vac}_\ell\times {\rm vac}_\ell) =
({\rm vac}_\ell \times {\rm vac}_\ell)^{\Omega=+}_+
+
({\rm vac}_\ell \times {\rm vac}_\ell)^{\Omega=-}_-
\;.
\]
This means the (anti)symmetric components in this tensor product have
positive (negative) parity.%
\footnote{Note that symmetrization is understood in a graded sense,
\ie products two fermions receive the opposite symmetry.}
This implies in particular, that all fermions in 
$({\rm vac}_\ell\times {\rm vac}_\ell)$, which need to be products of 
one fermion and one boson, always come in both parities $\Omega=\pm$.

\section*{Acknowledgements}

This work was supported in part by I.N.F.N.,
by the EC programs HPRN-CT-2000-00122, HPRN-CT-2000-00131 and
HPRN-CT-2000-00148, by the INTAS contract 99-1-590, by the MURST-COFIN
contract 2001-025492 and by the NATO contract PST.CLG.978785.
N.B. dankt der \emph{Studienstiftung des
deutschen Volkes} f\"ur die Unterst\"utzung durch ein 
Promotions\-f\"orderungsstipendium.

\appendix

\section*{Appendix}

\section{Notation of representations}
\label{app:notation}

In this appendix, we collect our notations for representations
of the various groups appearing in the main text. In general,
we denote representations ${\cal R}$ by their highest weight
states which are specified by their weights $(w_1, \dots w_n)$
or their Dynkin labels $[a_1, \dots, a_n]$.
For the groups $SO(4)$,
$SO(6)$, and $SO(10)$, used in the text, the change to the
$SU(2)_L\times SU(2)_R$, $SU(4)$ and $SO(10)$ Dynkin basis is
simply given as 
\< (s_1,s_2) \eq [s_1+s_2,s_1-s_2] \;, \nln{}
(j_1,j_2,j_3) \eq [j_2+j_3,j_1-j_2,j_2-j_3] \;, \nln{}
(w_1,w_2,w_3,w_4,w_5) \eq
[w_1-w_2,w_2-w_3,w_3-w_4,w_4+w_5,w_4-w_5] \;,\nln{}
 [s_1,s_2] \eq
(\half s_1+\half s_2,\half s_1-\half s_2) \;, 
\nln{} [q_1,p,q_2] \eq (p+\half q_1+\half
q_2,\half q_1+\half q_2,\half q_1-\half q_2)\;, \nln{}
[k,l,m,r_1,r_2] \eq (k+l+m+\half r_1+\half r_2,l+m+\half r_1+\half
r_2,m+\half r_1+\half r_2,
          \nl \qquad\qquad\half r_1+\half r_2,\half r_1-\half r_2)
\;,
 \>
 respectively .

By $\chi_{\wfat}(y_i)$, we denote the
character polynomial associated to the
highest weight representation 
${\cal R}_{\mathbf{w}}$ of $SO(6)$ or $SO(4)$:
 \[
 \chi_{\mathbf{w}}(y_i)\equiv 
 \sum_{\mathbf{w'}\in {\cal
R}_{\mathbf{w}}}\, \yfat^{\mathbf{w'}} \;,
\qquad
{\bf y}^{\bf w'}\equiv \prod_{i}\, y_i^{w'_i}
\;.
\]
The set of ${\bf w'}$ over which the sum runs may be
determined recursively, using Freudenthal's multiplicity
formula, or directly by use of Weyl's character formula, cf.\
Appendix~\ref{app:tools}.

E.g., for the first few irreps (${\bf 1}$, ${\bf 6}$, ${\bf 4}$,
and ${\bf 4^*}$) of $SO(6)$ one has
\<
 \chi_{(000)}\eq 1
\;,\nln
 \chi_{(100)}\eq y_1 +y_1^{-1}+y_2 +y_2^{-1}+y_3 +y_3^{-1}
\;,\nln
 \chi_{(\frac{1}{2},\frac{1}{2},\frac{1}{2})}\eq
   y_1^\frac{1}{2} y_2^\frac{1}{2} y_3^\frac{1}{2}
  +y_1^\frac{1}{2} y_2^{-\frac{1}{2}} y_3^{-\frac{1}{2}}
  +y_1^{-\frac{1}{2}} y_2^{-\frac{1}{2}} y_3^\frac{1}{2}
  +y_1^{-\frac{1}{2}} y_2^\frac{1}{2} y_3^{-\frac{1}{2}}
\;,\nln
 \chi_{(\frac{1}{2},\frac{1}{2},-\frac{1}{2})}\eq
   y_1^\frac{1}{2} y_2^\frac{1}{2} y_3^{-\frac{1}{2}}
  +y_1^\frac{1}{2} y_2^{-\frac{1}{2}} y_3^{\frac{1}{2}}
  +y_1^{-\frac{1}{2}} y_2^{\frac{1}{2}} y_3^\frac{1}{2}
  +y_1^{-\frac{1}{2}} y_2^{-\frac{1}{2}} y_3^{-\frac{1}{2}} \label{so6}\;,
\>
while a generic irrep of $SO(4)$ is given by
\<
 \chi_{(s_1,s_2)}\eq
  \frac{y_4^{-s_2} y_5^{-s_1} +
  y_4^{ s_2} y_5^{2+s_1}- y_4^{s_1+1} y_5^{s_2+1} -
    y_4^{-1-s_1} y_5^{1-s_2}}{(1-y_4 y_5)(1-y_4^{-1} y_5)}
    \; \label{so4}.
\>
Tensor products of representations ${\cal R},{\cal R}'$ translate
into ordinary product of their character
polynomials: $\chi_{\reps\times\reps'}=\chi_{\reps}\chi_{\reps'}$.

\section{String primaries}
\label{app:string}

In this appendix, we give a list of the first ten massive
flat space string levels, organized under 
$SO(10)\times SO(2)_\Delta$, as described in section~\ref{sec:string}. We use the notation 
\begin{eqnarray}
[k,l,m,p,q]^* &\equiv& [k,l,m,p,q]-[k\!-\!1,l,m,p,q] \;,
\qquad (k>0) \;.
\end{eqnarray}

\bigskip

     \noindent$\ell=1$ :
     {\footnotesize\begin{equation}\nonumber\begin{tabular}{l|l}
$\Delta$ & ${\cal R}$ \\ \hline 
     $1$ &   $
     [0, 0, 0, 0, 0]
     $\\ \hline
     \end{tabular}\end{equation}}
     \bigskip

     \noindent$\ell=2$ :
     {\footnotesize\begin{equation}\nonumber\begin{tabular}{l|l}
     $\Delta$ & ${\cal R}$ \\ \hline
     $2$ &   $
     [1, 0, 0, 0, 0]^*
     $\\ \hline
     \end{tabular}\end{equation}}
     \bigskip

     \noindent$\ell=3$ :
     {\footnotesize\begin{equation}\nonumber\begin{tabular}{l|l}
     $\Delta$ & ${\cal R}$ \\ \hline
     $3$ &   $
     [2, 0, 0, 0, 0]^*
     $\\ \hline
     $\frac{5}{2}$ &   $
     [0, 0, 0, 0, 1]
     $\\ \hline
     \end{tabular}\end{equation}}
     \bigskip

     \noindent$\ell=4$ :
     {\footnotesize\begin{equation}\nonumber\begin{tabular}{l|l}
     $\Delta$ & ${\cal R}$ \\ \hline
     $4$ &   $
     [3, 0, 0, 0, 0]^*
     $\\ \hline
     $\frac{7}{2}$ &   $
     [1, 0, 0, 0, 1]^*
     $\\ \hline
     $3$ &   $
     [0, 1, 0, 0, 0]
     $\\ \hline
     \end{tabular}\end{equation}}
     \bigskip

     \noindent$\ell=5$ :
     {\footnotesize\begin{equation}\nonumber\begin{tabular}{l|l}
     $\Delta$ & ${\cal R}$ \\ \hline
     $5$ &   $
     [4, 0, 0, 0, 0]^*
     $\\ \hline
     $\frac{9}{2}$ &   $
     [2, 0, 0, 0, 1]^*
     $\\ \hline
     $4$ &   $
     [0, 0, 1, 0, 0]
     +[1, 1, 0, 0, 0]^*
     $\\ \hline
     $\frac{7}{2}$ &   $
     [1, 0, 0, 0, 1]
     $\\ \hline
     $3$ &   $
     [0, 0, 0, 0, 0]
     $\\ \hline
     \end{tabular}\end{equation}}
     \bigskip

     \noindent$\ell=6$ :
     {\footnotesize\begin{equation}\nonumber\begin{tabular}{l|l}
     $\Delta$ & ${\cal R}$ \\ \hline
     $6$ &   $
     [5, 0, 0, 0, 0]^*
     $\\ \hline
     $\frac{11}{2}$ &   $
     [3, 0, 0, 0, 1]^*
     $\\ \hline
     $5$ &   $
     [1, 0, 1, 0, 0]^*
     +[2, 1, 0, 0, 0]^*
     $\\ \hline
     $\frac{9}{2}$ &   $
     [0, 1, 0, 0, 1]
     +[1, 0, 0, 1, 0]
     +[2, 0, 0, 0, 1]^*
     $\\ \hline
     $4$ &   $
     [0, 0, 0, 0, 2]
     +[1, 0, 0, 0, 0]^*
     +[1, 1, 0, 0, 0]
     $\\ \hline
     $\frac{7}{2}$ &   $
     [0, 0, 0, 1, 0]
     $\\ \hline
     \end{tabular}\end{equation}}
     \bigskip

     \noindent$\ell=7$ :
     {\footnotesize\begin{equation}\nonumber\begin{tabular}{l|l}
     $\Delta$ & ${\cal R}$ \\ \hline
     $7$ &   $
     [6, 0, 0, 0, 0]^*
     $\\ \hline
     $\frac{13}{2}$ &   $
     [4, 0, 0, 0, 1]^*
     $\\ \hline
     $6$ &   $
     [2, 0, 1, 0, 0]^*
     +[3, 1, 0, 0, 0]^*
     $\\ \hline
     $\frac{11}{2}$ &   $
     [0, 1, 0, 1, 0]
     +[1, 1, 0, 0, 1]^*
     +[2, 0, 0, 1, 0]^*
     +[3, 0, 0, 0, 1]^*
     $\\ \hline
     $5$ &   $
     [0, 0, 0, 1, 1]
     +[0, 2, 0, 0, 0]
     +[1, 0, 0, 0, 2]^*
     +[1, 0, 1, 0, 0]
     $\\ &   $
     +[2, 0, 0, 0, 0]+[2,0,0,0,0]^*
     +[2, 1, 0, 0, 0]^*
     $\\ \hline
     $\frac{9}{2}$ &   $
     [0, 0, 0, 0, 1]
     +[0, 1, 0, 0, 1]
     +[1, 0, 0, 1, 0]^*
     +[2, 0, 0, 0, 1]
     $\\ \hline
     $4$ &   $
     [0, 0, 1, 0, 0]
     +[1, 0, 0, 0, 0]
     $\\ \hline
     \end{tabular}\end{equation}}
     \bigskip

     \noindent$\ell=8$ :
     {\footnotesize\begin{equation}\nonumber\begin{tabular}{l|l}
     $\Delta$ & ${\cal R}$ \\ \hline
     $8$ &   $
     [7, 0, 0, 0, 0]^*
     $\\ \hline
     $\frac{15}{2}$ &   $
     [5, 0, 0, 0, 1]^*
     $\\ \hline
     $7$ &   $
     [3, 0, 1, 0, 0]^*
     +[4, 1, 0, 0, 0]^*
     $\\ \hline
     $\frac{13}{2}$ &   $
     [1, 1, 0, 1, 0]^*
     +[2, 1, 0, 0, 1]^*
     +[3, 0, 0, 1, 0]^*
     +[4, 0, 0, 0, 1]^*
     $\\ \hline
     $6$ &   $
     [0, 0, 0, 2, 0]
     +[0, 1, 1, 0, 0]
     +[1, 0, 0, 1, 1]^*
     +[1, 1, 0, 0, 0]
     +[1, 2, 0, 0, 0]^*
     $ \\& $
     +[2, 0, 0, 0, 2]^*
     +[2, 0, 1, 0, 0]^*
     +2 \!\cdot\! [3, 0, 0, 0, 0]^*
     +[3, 1, 0, 0, 0]^*
     $\\ \hline
     $\frac{11}{2}$ &   $
     [0, 0, 1, 0, 1]
     +[0, 1, 0, 1, 0]
     +[1, 0, 0, 0, 1]+[1,0,0,0,1]^*
     +[1, 1, 0, 0, 1]+[1,1,0,0,1]^*
     $ \\& $
     +[2, 0, 0, 1, 0]+[2,0,0,1,0]^*
     +[3, 0, 0, 0, 1]^*
     $\\ \hline
     $5$ &   $
     [0, 0, 0, 1, 1]
     +2 \!\cdot\! [0, 1, 0, 0, 0]
     +[1, 0, 0, 0, 2]
     +[1, 0, 1, 0, 0]+[1,0,1,0,0]^*
$\\ 
     &$     
     +[2, 0, 0, 0, 0]^* 
     +[2, 1, 0, 0, 0]
     $\\ \hline
     $\frac{9}{2}$ &   $
     [0, 0, 0, 0, 1]
     +[0, 1, 0, 0, 1]
     +[1, 0, 0, 1, 0]
     $\\ \hline
     $4$ &   $
     [1, 0, 0, 0, 0]
     $\\ \hline
     \end{tabular}\end{equation}}
     \bigskip

     \noindent$\ell=9$ :
     {\footnotesize\begin{equation}\nonumber\begin{tabular}{l|l}
     $\Delta$ & ${\cal R}$ \\ \hline
     $9$ &   $
     [8, 0, 0, 0, 0]^*
     $\\ \hline
     $\frac{17}{2}$ &   $
     [6, 0, 0, 0, 1]^*
     $\\ \hline
     $8$ &   $
     [4, 0, 1, 0, 0]^*
     +[5, 1, 0, 0, 0]^*
     $\\ \hline
     $\frac{15}{2}$ &   $
     [2, 1, 0, 1, 0]^*
     +[3, 1, 0, 0, 1]^*
     +[4, 0, 0, 1, 0]^*
     +[5, 0, 0, 0, 1]^*
     $\\ \hline
     $7$ &   $
     [0, 2, 0, 0, 0]
     +[1, 0, 0, 2, 0]^*
     +[1, 1, 1, 0, 0]^*
     +[2, 0, 0, 1, 1]^*
     +[2, 1, 0, 0, 0]^*
     $ \\& $
     +[2, 2, 0, 0, 0]^*
     +[3, 0, 0, 0, 2]^*
     +[3, 0, 1, 0, 0]^*
     +2 \!\cdot\! [4, 0, 0, 0, 0]^*
     +[4, 1, 0, 0, 0]^*
     $\\ \hline
     $\frac{13}{2}$ &   $
     [0, 0, 1, 1, 0]
     +[0, 1, 0, 0, 1]
     +[0, 2, 0, 0, 1]
     +[1, 0, 0, 1, 0]
     +[1, 0, 1, 0, 1]^*
     +[1, 1, 0, 1, 0]     $ \\& $
+[1,1,0,1,0]^*
     +2 \!\cdot\! [2, 0, 0, 0, 1]^*
     +2 \!\cdot\! [2, 1, 0, 0, 1]^*
     +2 \!\cdot\! [3, 0, 0, 1, 0]^*
     +[4, 0, 0, 0, 1]^*
     $\\ \hline
     $6$ &   $
     2 \!\cdot\! [0, 0, 1, 0, 0]
     +[0, 1, 0, 0, 2]
     +[0, 1, 1, 0, 0]
     +2 \!\cdot\! [1, 0, 0, 1, 1]+[1,0,0,1,1]^*
     $ \\& $
     +[1, 1, 0, 0, 0]+2 \!\cdot\! [1, 1, 0, 0, 0]^*
     +[1, 2, 0, 0, 0]
     +[2, 0, 0, 0, 2]^*
     +[2, 0, 1, 0, 0]     $ \\& $
+2 \!\cdot\! [2, 0, 1, 0, 0]^*
     +[3, 0, 0, 0, 0]+[3,0,0,0,0]^*
     +[3, 1, 0, 0, 0]^*
     $\\ \hline
     $\frac{11}{2}$ &   $
     [0, 0, 0, 1, 0]
     +2 \!\cdot\! [0, 0, 1, 0, 1]
     +2 \!\cdot\! [0, 1, 0, 1, 0]
     +2 \!\cdot\! [1, 0, 0, 0, 1]+[1, 0, 0, 0, 1]^*
     $ \\& $
     +[1, 1, 0, 0, 1]+[1,1,0,0,1]^*
     +[2, 0, 0, 1, 0]+[2,0,0,1,0]^*
     +[3, 0, 0, 0, 1]
     $\\ \hline
     $5$ &   $
     [0, 0, 0, 0, 0]
     +[0, 0, 0, 1, 1]
     +2 \!\cdot\! [0, 1, 0, 0, 0]
     +[0, 2, 0, 0, 0]
     +[1, 0, 0, 0, 2]
     +[1, 0, 1, 0, 0]
     $ \\& $
     +[2, 0, 0, 0, 0]+[2,0,0,0,0]^*
     $\\ \hline
     $\frac{9}{2}$ &   $
     [0, 0, 0, 0, 1]
     +[1, 0, 0, 1, 0]
     $\\ \hline
     \end{tabular}\end{equation}}
     \bigskip

     \noindent$\ell=10$ :
     {\footnotesize\begin{equation}\nonumber\begin{tabular}{l|l}
     $\Delta$ & ${\cal R}$ \\ \hline
     $10$ &   $
     [9, 0, 0, 0, 0]^*
     $\\ \hline
     $\frac{19}{2}$ &   $
     [7, 0, 0, 0, 1]^*
     $\\ \hline
     $9$ &   $
     [5, 0, 1, 0, 0]^*
     +[6, 1, 0, 0, 0]^*
     $\\ \hline
     $\frac{17}{2}$ &   $
     [3, 1, 0, 1, 0]^*
     +[4, 1, 0, 0, 1]^*
     +[5, 0, 0, 1, 0]^*
     +[6, 0, 0, 0, 1]^*
     $\\ \hline
     $8$ &   $
     [1, 2, 0, 0, 0]^*
     +[2, 0, 0, 2, 0]^*
     +[2, 1, 1, 0, 0]^*
     +[3, 0, 0, 1, 1]^*
     +[3, 1, 0, 0, 0]^*
     $ \\& $
     +[3, 2, 0, 0, 0]^*
     +[4, 0, 0, 0, 2]^*
     +[4, 0, 1, 0, 0]^*
     +2 \!\cdot\! [5, 0, 0, 0, 0]^*
     +[5, 1, 0, 0, 0]^*
     $\\ \hline
     $\frac{15}{2}$ &   $
     [0, 1, 0, 1, 0]
     +[0, 2, 0, 1, 0]
     +[1, 0, 1, 1, 0]^*
     +[1, 1, 0, 0, 1]^*
     +[1, 2, 0, 0, 1]^*
     $ \\& $
     +[2, 0, 0, 1, 0]^*
     +[2, 0, 1, 0, 1]^*
     +2 \!\cdot\! [2, 1, 0, 1, 0]^*
     +2 \!\cdot\! [3, 0, 0, 0, 1]^*
     +2 \!\cdot\! [3, 1, 0, 0, 1]^*
     $ \\& $
     +2 \!\cdot\! [4, 0, 0, 1, 0]^*
     +[5, 0, 0, 0, 1]^*
     $\\ \hline
     $7$ &   $
     [0, 0, 0, 1, 1]
     +[0, 1, 0, 0, 0]
     +2 \!\cdot\! [0, 1, 0, 1, 1]
     +[0, 2, 0, 0, 0]
     +[0, 3, 0, 0, 0]
     +[1, 0, 0, 2, 0]
     $ \\& $
     +[1, 0, 1, 0, 0]+2 \!\cdot\! [1, 0, 1, 0, 0]^*
     +[1, 1, 0, 0, 2]^*
     +[1, 1, 1, 0, 0]+[1,1,1,0,0]^*
     $ \\& $
     +3 \!\cdot\! [2, 0, 0, 1, 1]^*
     +[2, 1, 0, 0, 0]+3 \!\cdot\! [2, 1, 0, 0, 0]^*
     +[2, 2, 0, 0, 0]^*
     +[3, 0, 0, 0, 2]^*
     $ \\& $
     +3 \!\cdot\! [3, 0, 1, 0, 0]^*
     +2 \!\cdot\! [4, 0, 0, 0, 0]^*
     +[4, 1, 0, 0, 0]^*
     $\\ \hline
     $\frac{13}{2}$ &   $
     [0, 0, 0, 1, 2]
     +2 \!\cdot\! [0, 0, 1, 1, 0]
     +3 \!\cdot\! [0, 1, 0, 0, 1]
     +[0, 2, 0, 0, 1]
     +2 \!\cdot\! [1, 0, 0, 1, 0]     $ \\& $
+[1, 0, 0, 1, 0]^*
     +[1, 0, 1, 0, 1]+2 \!\cdot\! [1, 0, 1, 0, 1]^*
     +2 \!\cdot\! [1, 1, 0, 1, 0]+2 \!\cdot\! [1, 1, 0, 1, 0]^*
     $ \\& $
     +2 \!\cdot\! [2, 0, 0, 0, 1]+3 \!\cdot\! [2, 0, 0, 0, 1]^*
     +[2, 1, 0, 0, 1]+2 \!\cdot\! [2, 1, 0, 0, 1]^*
     +[3, 0, 0, 1, 0]+     $ \\& $
2 \!\cdot\! [3, 0, 0, 1, 0]^*
     +[4, 0, 0, 0, 1]^*
     $\\ \hline
     $6$ &   $
     2 \!\cdot\! [0, 0, 0, 0, 2]
     +[0, 0, 0, 2, 0]
     +2 \!\cdot\! [0, 0, 1, 0, 0]
     +[0, 1, 0, 0, 2]
     +3 \!\cdot\! [0, 1, 1, 0, 0]
     $ \\& $
     + [1, 0, 0, 0, 0]+ [1, 0, 0, 0, 0]^*
     +3 \!\cdot\! [1, 0, 0, 1, 1]+[1, 0, 0, 1, 1]^*
     +3 \!\cdot\! [1, 1, 0, 0, 0]+     $ \\& $
2 \!\cdot\! [1, 1, 0, 0, 0]^*
     +[1, 2, 0, 0, 0]^*
     + [2, 0, 0, 0, 2]+ [2, 0, 0, 0, 2]^*
     +[2, 0, 1, 0, 0]          $ \\& $
 + [2, 0, 1, 0, 0]^*
    + [3, 0, 0, 0, 0]+2 \!\cdot\! [3, 0, 0, 0, 0]^*
     +[3, 1, 0, 0, 0]
     $\\ \hline
     $\frac{11}{2}$ &   $
     [0, 0, 0, 0, 3]
     +2 \!\cdot\! [0, 0, 0, 1, 0]
     +[0, 0, 1, 0, 1]
     +2 \!\cdot\! [0, 1, 0, 1, 0]
     +3 \!\cdot\! [1, 0, 0, 0, 1]     $ \\& $
+[1, 0, 0, 0, 1]^*
     +2 \!\cdot\! [1, 1, 0, 0, 1]
     + [2, 0, 0, 1, 0]+ [2, 0, 0, 1, 0]^*
     $\\ \hline
     $5$ &   $
     [0, 0, 0, 1, 1]
     +2 \!\cdot\! [0, 1, 0, 0, 0]
     +[1, 0, 1, 0, 0]
     +[2, 0, 0, 0, 0]
     $\\ \hline
     $\frac{9}{2}$ &   $
     [0, 0, 0, 0, 1]
     $\\ \hline
     \end{tabular}\end{equation}}
     \bigskip

\section{String partition function}
\label{app:stringpart}


In this appendix, we evaluate the string partition
function
 \[
 \mathcal{Z}(q_1,q_2)~=~\Tr \left[(-)^F \, q_1^{\ell-\bar\ell}\,
q_2^{\nu+\bar{\nu}} \right] \;,
 \]
with $q_1=e^{2\pi i \tau_1}$,\quad $q_2=e^{-2\pi \tau_2}$,\quad
$\ell=L_0-\frac{c}{ 24}=\sum_n n N_n$,\quad
$\bar{\ell}=\bar{L}_0-\frac{c}{ 24}=\sum_n n \bar{N}_n$,\quad
$\nu=\sum_n N_n w_n$,\quad $\bar{\nu}=\sum_n \bar{N}_n w_n$ the
left-right moving string levels and excitation numbers corresponding
to the chiral contributions to the worldsheet momentum and
Hamiltonian, respectively.  Physical states are identified by the
level matching condition $\ell=\bar{\ell}$.  We start by considering
the GS string in flat space ($w_n=n$):
\begin{eqnarray}
\mathcal{Z}_{\rm flat}(q_1,q_2) &=& ({\bf 8}_v-{\bf 8_c})^2
\prod_{n=1}^\infty 
\frac{ (1- q_1^n q_2^n)^{\bf 8_s}(1- q_1^{-n} q_2^n)^{\bf 8_s} }
{ (1- q_1^{n} q_2^n)^{\bf 8_v}(1- q_1^{-n} q_2^n)^{\bf 8_v}  }
 \;,
\label{zflat}
 \end{eqnarray}
with
 \begin{eqnarray}
  (1-q^n)^{\bf 8_s} &=&1-{\bf 8_s}\, q^n 
+ {\bf 8_s}\wedge {\bf 8_s}\, q^{2n}+\ldots  \nn\\
  (1-q^n )^{-{\bf 8_v}}&=& 1+{\bf 8_v}\, 
q^n+ {\bf 8_{(v}}\times {\bf 8_{v)}}\, q^{2n} +\ldots
\end{eqnarray}
As argued in the main text, the AdS string partition function can be
found from that of string theory in flat space after replacing string
frequencies $w_n$ by those of strings on plane wave at $\gym=0$ \ie
$w_n\to 1$ and extending the product in (\ref{zflat}) to $n=0$ to
account for the new zero modes:
\begin{eqnarray}
 q_2^{n}  & \to&  q_2\nn\;,\qquad
({\bf 8}_v-{\bf 8_c})^2 ~\to~  
\frac{(1-q_2)^{\bf 8_c}}{(1- q_2)^{\bf 8_v}} \;. 
\label{pprep}
\end{eqnarray}
We are interested in massive string states $n\geq 1$. Primaries are
found by dividing by $({\bf 8}_v-{\bf 8}_s)^2$ and suppressing from
bosonic and fermionic zero modes. Restricting to the chiral partition
function one finds:
  \begin{eqnarray}
\mathcal{Z}_{\rm SO(9)}(q_1,q_2) &\equiv& 
\sum_{\ell,\nu} d_{\ell,\nu}\, q_1^\ell q_2^\nu
= \frac{1}{ ({\bf 8}_v-{\bf 8}_s)}\, 
\left[ \prod_{n=1}^\infty \frac{ (1- q_1^n q_2)^{\bf 8_s} }{ (1-q_1^n q_2)^{\bf 8_v}  }-1\right]\nn\\[1ex]
&=&  q_2\,q_1 + \left( q_2 + 8\,q_2^2 \right) \,q_1^2  +  \left(   -16\,q_2^2 +q_2+ 43\,q_2^3 \right)
\,   q_1^3 + \ldots \;.  \label{qu}
  \label{so9u}
 \end{eqnarray}
This is to be compared with the third line in (\ref{second}), e.g. at
$\ell=2$ i.e.\ $q_1^2$, we find $q_2 + 8\,q_2^2 \leftrightarrow
[0000]_1+[1000]_2$. Expanding (\ref{so9u}) one can easily extend this
result to higher string levels.  At $q_2=1$ states organized in
$SO(9)$ representations as expected. Finally the lift to $SO(10)$ can
be implemented by multiplying and dividing by $(1-q_1 q_2^2)$
\begin{eqnarray}
\mathcal{Z}_{\rm SO(10)}(q_1,q_2) &=&  \frac{(1-q_1 q_2^2)}{ (1-q_1 q_2^2)}\,
\mathcal{Z}_{\rm SO(9)}(q_1,q_2)\nn\\
&=& \frac{1}{ ({\bf 8}_v-{\bf 8}_s)}\,\frac{(1-q_1 q_2^2)}{ (1-q_1 q_2^2)}
\left[ \prod_{n=1}^\infty \frac{ (1- q_1^n q_2)^{\bf 8_s}}{(1- q_1^n q_2)^{\bf 8_v}  }-1\right] 
\;, \label{so10u}
\end{eqnarray}
\ie~states at level $q_1^\ell$ are added and subtracted at level
$q_1^{\ell+1}$ \eg at level $\ell=2$, we have the lift
 $q_2 + 8\,q_2^2 \to q_2 + 8\,q_2^2+q_2^3-q_2^3$ in agreement
with the second line in (\ref{second}) with the power $\nu$ of $q_2$
given as $\nu=\Delta-J$. 
This completes $SO(10)$ representations  as can be
easily seen by expanding (\ref{so10u}) and setting $q_2\to 1$ (keeping 
states with negative multiplicities). The
string partition function (\ref{so10u})
 can be used to read off multiplicities and $SO(8)\times
SO(2)_{\Delta-J}$ charges. The assignments of conformal
dimensions $\Delta$ to string states then follows from the
requirement that $SO(10)$ string representations reduce
consistently to (\ref{so9u}) via (\ref{magic}).


\section{Free spectrum of $\superN=4$ SYM}
\label{app:specN4}

In this appendix we present the spectrum of long multiplets in
$\superN=4$ SYM in terms of its superconformal primaries, \ie
$\zprim(t,y_i)$.  We have computed this spectrum up to $\Delta=10$,
but for reasons of limited space, here we give the result only up to
$\Delta=\frac{13}{2}$.

A long multiplet of $\superN=4$ is described by the symbol
\[[s_1,s_2;q_1,p,q_2]_{L,B}^P \;.\]
Here, $[s_1,s_2]$ and $[q_1,p,q_2]$ indicate the
Dynkin labels of $SO(4)$ and $SO(6)$, respectively.
In addition we write the
parity $P$, hypercharge $B$ and length $L$ of a state.
Parity is described in \cite{Beisert:2003jj},
the hypercharge $B$ corresponds to the leading order $U(1)_B$
charge in the decomposition
$SU(2,2|4)=U(1)_B\ltimes PSU(2,2|4)$,
and the length $L$ is the leading order number of letters used to
construct the state.
Furthermore, $P=\pm$ indicates a pair with opposite parities
and $+\mathrm{conj.}$ indicates a conjugate
state $[s_2,s_1;q_2,p,q_1]_{L,-B}^P$.
We group the multiplets according to their 
classical dimension $\Delta$.

{\footnotesize\[\nonumber\begin{array}{l|l}
\Delta&\mathcal{R}\\
\hline
2&
[0,0;0,0,0]_{2,0}^{+}
\\\hline
3&
[0,0;0,1,0]_{3,0}^{-}
\\\hline
4&
2\mathord{\cdot}[4;0,0;0,0,0]_{4,0}^{+}
+[0,0;1,0,1]_{4,0}^{-}
+2\mathord{\cdot}[4;0,0;0,2,0]_{4,0}^{+}
\\&\mathord{}
+([0,2;0,0,0]_{3,-1}^{-}+\mathrm{conj.})
+[1,1;0,1,0]_{3,0}^{\pm}
+[2,2;0,0,0]_{2,0}^{+}
\\\hline
5&
4\mathord{\cdot}[0,0;0,1,0]_{5,0}^{-}
+2\mathord{\cdot}([0,0;0,0,2]_{5,0}^{+}+\mathrm{conj.})
+[0,0;1,1,1]_{5,0}^{\pm}
\\&\mathord{}
+2\mathord{\cdot}[5;0,0;0,3,0]_{5,0}^{-}
+2\mathord{\cdot}([0,2;0,1,0]_{4,-1}^{+}+\mathrm{conj.})
\\&\mathord{}
+([0,2;2,0,0]_{4,-1}^{-}+\mathrm{conj.})
+[1,1;0,0,0]_{4,0}^{\pm}
+2\mathord{\cdot}[1,1;1,0,1]_{4,0}^{\pm}
\\&\mathord{}
+[1,1;0,2,0]_{4,0}^{\pm}
+[2,2;0,1,0]_{3,0}^{-}
\\\hline
\sfrac{11}{2}&
2\mathord{\cdot}[0,1;1,0,0]_{5,-1/2}^{\pm}
+2\mathord{\cdot}[0,1;0,1,1]_{5,-1/2}^{\pm}
+[0,1;1,2,0]_{5,-1/2}^{\pm}
\\&\mathord{}
+2\mathord{\cdot}[1,2;0,0,1]_{4,-1/2}^{\pm}
+[1,2;1,1,0]_{4,-1/2}^{\pm}
+[2,3;1,0,0]_{3,-1/2}^{\pm}
+\mathrm{conjugates}
\\\hline
6&
2\mathord{\cdot}[0,0;0,0,0]_{4,0}^{+}
+2\mathord{\cdot}([0,0;0,0,0]_{5,-1}^{+}+\mathrm{conj.})
+5\mathord{\cdot}[0,0;0,0,0]_{6,0}^{+}
+3\mathord{\cdot}[0,0;1,0,1]_{6,0}^{+}
\\&\mathord{}
+6\mathord{\cdot}[0,0;1,0,1]_{6,0}^{-}
+9\mathord{\cdot}[0,0;0,2,0]_{6,0}^{+}
+[0,0;0,2,0]_{6,0}^{-}
+3\mathord{\cdot}([0,0;0,1,2]_{6,0}^{-}+\mathrm{conj.})
\\&\mathord{}
+3\mathord{\cdot}[0,0;2,0,2]_{6,0}^{+}
+[0,0;1,2,1]_{6,0}^{+}
+2\mathord{\cdot}[0,0;1,2,1]_{6,0}^{-}
+3\mathord{\cdot}[0,0;0,4,0]_{6,0}^{+}
\\&\mathord{}
+2\mathord{\cdot}([0,2;0,0,0]_{4,0}^{-}+\mathrm{conj.})
+3\mathord{\cdot}([0,2;0,0,0]_{5,-1}^{-}+\mathrm{conj.})
\\&\mathord{}
+4\mathord{\cdot}([0,2;1,0,1]_{5,-1}^{+}+\mathrm{conj.})
+2\mathord{\cdot}([0,2;1,0,1]_{5,-1}^{-}+\mathrm{conj.})
\\&\mathord{}
+4\mathord{\cdot}([0,2;0,2,0]_{5,-1}^{-}+\mathrm{conj.})
+([0,2;2,1,0]_{5,-1}^{+}+\mathrm{conj.})
+8\mathord{\cdot}[1,1;0,1,0]_{5,0}^{\pm}
\\&\mathord{}
+2\mathord{\cdot}([1,1;0,0,2]_{5,0}^{\pm}+\mathrm{conj.})
+4\mathord{\cdot}[1,1;1,1,1]_{5,0}^{\pm}
+2\mathord{\cdot}[1,1;0,3,0]_{5,0}^{\pm}
\\&\mathord{}
+([0,4;0,0,0]_{3,-1}^{+}+\mathrm{conj.})
+([0,4;0,0,0]_{4,-2}^{+}+\mathrm{conj.})
+2\mathord{\cdot}([1,3;0,1,0]_{4,-1}^{\pm}+\mathrm{conj.})
\\&\mathord{}
+5\mathord{\cdot}[2,2;0,0,0]_{4,0}^{+}
+2\mathord{\cdot}[2,2;0,0,0]_{4,0}^{-}
+2\mathord{\cdot}[2,2;1,0,1]_{4,0}^{-}
+4\mathord{\cdot}[2,2;0,2,0]_{4,0}^{+}
\\&\mathord{}
+[2,2;0,2,0]_{4,0}^{-}
+([2,4;0,0,0]_{3,-1}^{-}+\mathrm{conj.})
+[3,3;0,1,0]_{3,0}^{\pm}
+[4,4;0,0,0]_{2,0}^{+}
\\\hline
\frac{13}{2}&
4\mathord{\cdot}[0,1;0,0,1]_{5,+1/2}^{\pm}
+6\mathord{\cdot}[0,1;0,0,1]_{6,-1/2}^{\pm}
+12\mathord{\cdot}[0,1;1,1,0]_{6,-1/2}^{\pm}
\\&\mathord{}
+5\mathord{\cdot}[0,1;1,0,2]_{6,-1/2}^{\pm}
+[0,1;3,0,0]_{6,-1/2}^{\pm}
+5\mathord{\cdot}[0,1;0,2,1]_{6,-1/2}^{\pm}
\\&\mathord{}
+2\mathord{\cdot}[0,1;2,1,1]_{6,-1/2}^{\pm}
+[0,1;1,3,0]_{6,-1/2}^{\pm}
+[0,3;0,0,1]_{4,-1/2}^{\pm}
\\&\mathord{}
+[0,3;0,0,1]_{5,-3/2}^{\pm}
+2\mathord{\cdot}[0,3;1,1,0]_{5,-3/2}^{\pm}
+10\mathord{\cdot}[1,2;1,0,0]_{5,-1/2}^{\pm}
\\&\mathord{}
+8\mathord{\cdot}[1,2;0,1,1]_{5,-1/2}^{\pm}
+3\mathord{\cdot}[1,2;2,0,1]_{5,-1/2}^{\pm}
+2\mathord{\cdot}[1,2;1,2,0]_{5,-1/2}^{\pm}
\\&\mathord{}
+[1,4;1,0,0]_{4,-3/2}^{\pm}
+3\mathord{\cdot}[2,3;0,0,1]_{4,-1/2}^{\pm}
+2\mathord{\cdot}[2,3;1,1,0]_{4,-1/2}^{\pm}
+\mathrm{conjugates}\\
\hline
\end{array}
\]}

\section{Tools for Representations}
\label{app:tools}

In this appendix we present two algorithms to
construct and deconstruct multiplets of
$SO(2n)$.

\subsection{Construction}

The character polynomial $\chi_{\bf w}$
associated with the highest weight state $\yfat^{\bf w}$ may
be directly obtained from Weyl's character formula. For
$SO(2n)$, this takes the form
\[
\chi_{\bf w} =
\frac{\sum_W (-)^{|W|}\,   W( \yfat^{\bf w+\rho} )}
{\sum_W (-)^{|W|} \,  W( \yfat^{\rho})} \;,
\]
where $\rho=(n\!-\!1,n\!-\!2,\dots,0)$ denotes one half the
sum of the positive roots. The sums are running over the Weyl
group of $SO(2n)$, which is generated by the group $S_n$ of
permutations of the $y_n$, together with the element
$\{ y_1 \to y_1^{-1}\,,\; y_2 \to y_2^{-1} \}\;$.

Another way to construct a multiplet is as follows.
We first decompose a highest weight multiplet
of $SO(2n)$ into multiplets of $SO(2n-2)\times SO(2)$.
This can be used to recursively construct all the weights in a
highest weight multiplet of $SO(2n)$.
The recursion formula is
\[
\chi_{(w_1,\ldots,w_{n})}=
\sum_{w'_i}\chi_{(w'_1,\ldots,w'_{n-1})}y_n^{c_n}\prod_{i=1}^{n-1}
\chi_{(c_i)}  \;,\]
with $\chi_{(c_i)}$ given by
\[
\chi_{(c_i)}=\frac{(y_n^{-c_i}-y_n^{c_i+2})}{(1-y_n^2)}
\;.
\]

The coefficients $c_i$ determining the range of
$SO(2)$ charges are given by
\<
c_1\eq w_1-\max (w'_1,w_{2})\geq 0,\nln
\earel{\vdots}\nln
c_i\eq\min (w_i,w_{i-1}')-\max (w'_i,w_{i+1})\geq 0,\nln
\earel{\vdots}\nln
c_{n-1}\eq\min (w_{n-1},w_{n-2}')-\max (|w'_{n-1}|,|w_{n}|)\geq0,\nln
c_n\eq\sign w_n\, \sign w'_{n-1}\, \min(|w_n|,|w'_{n-1}|),
\>
and the sum runs over those values of $w'_i$ for which the
coefficients $c_1,\ldots, c_{n-1}$ are non-negative.

\subsection{Deconstruction}

The highest weight $\wfat$ of a
irreducible multiplet $\chi_{\wfat}(y_i)$ of $SO(2n)$, $i=1,\ldots,n$  can
be obtained by a simple algorithm (alternating dominant) that yields
\[\label{yq}
\chi_{\wfat}\mapsto (\chi_{\wfat})\indup{HW}=\yfat^{\wfat}.
\]

We start by defining the fundamental Weyl chamber for a state
$\yfat^{(w_1,\ldots,w_n)}$
of $SO(2n)$ by the condition
\[\label{eq:FundWeyl}
w_1\geq w_2\geq\ldots\geq w_{n-1}\geq |w_n|.
\]
The algorithm moves all charges to the fundamental chamber by Weyl reflections.
These interchange two $SO(2)$ charges $w_k,w_l$ in the following way
\<
y_k^{w_k}\, y_l^{w_l} \earel{\mapsto } -y_k^{w_l+k-l}\, y_l^{w_k+l-k}\quad\quad\quad\quad~~~~~
 {\rm for}~~ k\neq l,
\nln
y_k^{w_k}\, y_n^{w_n} \earel{\mapsto}  -y_k^{-w_n+k-n}\, y_n^{-w_k+k-n}.
\>
Charges at the boundaries of chambers,
$w_l=w_k-k+l$ or $w_n=-w_k+k-n$, are dropped,
$y_k^{w_k}\, y_l^{w_k-k+l}\mapsto 0,y_k^{w_k}\, y_n^{-w_k+k+n}\mapsto 0 $.

When all the charges have been reflected to the fundamental chamber,
only one component, $\yfat^{\wfat}$, the highest weight state, should remain.
All the other components will have canceled among themselves.
As this algorithm is linear, it can equally well be used to decompose
any reducible representation of $SO(2n)$ into the highest weights of the
irreducible components
\[
\lrbrk{\sum_k \chi_{\wfat_k}}\indup{HW}=
\sum_k \yfat^{\wfat_k}.
\]
Especially, this algorithm can be applied to decompose the ${\cal N}=4$
partition function \eqref{polyah} into irreducible representations
of $SO(6)\times SO(4)$
\[
\zsym(t,y_i)\mapsto \zsym(t,y_i)\indup{HW}
\]
To this end it is particularly useful
 the following property (Klimyk's formula):
\[\label{eq:fasttensor}
(\chi \chi')\indup{HW}= (\chi\indup{HW} \chi')\indup{HW}.
\]
that allow us to determine the tensor product of two
multiplets, without computing the full product
of their character polynomials.
In other words, to determine the tensor product, we only need to compute the product of
the highest weights of one multiplet with all the weights of the other multiplet.
In our notation:
\[
\lrbrk{\chi_{\wfat}\chi_{\wfat'}}\indup{HW}=
\lrbrk{\yfat^{\wfat}\chi_{\wfat'}}\indup{HW}
\]

This can be used to significantly simplify expression \eqref{mlong} when we ask only for highest weights
with respect to $SO(6)\times SO(4)$.
According to \eqref{eq:fasttensor} and \eqref{yq}
only the highest weight ${\bf y}^{\wfat}$ enter in this reduction
and there is no need to construct the full multiplets of
$SO(4)\times SO(6)$.
 The result can then be read as:
 \[
 \chi_{\wfat}^{\Delta}
 (t,y_i)\indup{HW6\times 4}=
\left(\tlong(t,y_i)\, t^{\Delta}\,\chi_{\wfat}\right)\indup{HW6\times 4}=
\left(\tlong(t,y_i)\, t^{\Delta}\,\yfat^{\wfat}\right)\indup{HW6\times 4}.
 \label{mlonghw}
 \]
In a similar way one proceeds for BPS or semishort multiplets, with
final result again \eqref{mlonghw} but with the product over $s$
restricted to a subset of the supercharges $\wfat_s$.  For example the
highest weight content of a $\half$-BPS multiplet can be written as:
\[
\chi^n_{(n00;00)}(t,y_i)\indup{HW6\times 4}=
\left(t^n y_1^n\, \frac{\prod_{s=1}^{8}(1-t^\frac{1}{2}
\yfat^{\wfat_{s}})}{
 \prod_{v=1}^{4}(1-t\, \yfat^{\wfat_{v}})}\right)\indup{HW6\times 4}
\]
The product over supercharges is restricted to those $\wfat_{s}$ in
\eqref{weights} with $w_1=-\half$ \ie the unbroken supercharges.
Notice that expressions inside brackets $()\indup{HW}$ typically
involve incomplete multiplets of $SO(4)\times SO(6)$. It is
somewhat remarkable that, after removing negative weights, the
correct highest weights are produced by this expression.

Finally the algorithm provides an unambiguous procedure to
flip between irreps of
$SO(6)\times SO(4)$ and $SO(10)$ when the $SO(6)\times SO(4)$ representation
admits such a lift.
For that we identify states of $SO(10)$ and $SO(4)\times SO(6)$ by
$\yfat_{(w_1,w_2,w_3,w_4,w_5)}\equiv\yfat_{(w_1,w_2,w_3)}\yfat_{(w_4,w_5)}$.
To decompose a multiplet $\chi_{6\times 4}$
of $SO(6)\times SO(4)$ into highest weight states of
$SO(10)$ or vice versa we use
\[
\bigbrk{\chi_{6\times 4}}\indup{HW10}\qquad\mbox{or}\qquad
\bigbrk{\chi_{10}}\indup{HW6\times 4}.
\]
This can be applied to $\zten(t,y_i)$ in (\ref{zso10}) 
to read the $SO(10)$ content of SYM primaries  up to KK recurrences
$\zten(t,y_i)\indup{HW10}$.

\section{Cyclic Words}
\label{app:cyclic}

In this appendix we present a complete set of representatives for
cyclic words.
 We construct the weights of all single trace
operators in $\superN=4$ SYM. The module $\chi_L$ of all such
cyclic words of length $L$ is constructed as
\[
\chi_L=\sum_{W_{i}\in \chi\indup{F}} \Tr (W_{1}\ldots W_{L}).
\]
Each letter $W_{i}$ is summed over all fundamental fields in
\[
\chi\indup{F}=\{ \partial^s \phi, \partial^s \lambda ,\partial^s
\bar\lambda,\partial^s F\}
\]
Naively, we overcount due to cyclicity of traces.
Therefore we define $\Tr$ as to select exactly one representative
of each cyclic class. This can be done as follows
\[
\Tr (W_1\ldots W_L)=0\quad
\mbox{unless}\quad(W_1\ldots W_L)\leq
(W_{i+1}\ldots W_LW_1\ldots W_i)\mbox{ for all }i,
\]
where we define some ordering on the set of words.
Furthermore we need to determine whether the operator can exist at all
due to statistics. If we can write
$\Tr (W_1\ldots W_L)$ as $\Tr [(W_1\ldots W_{L/2})^{2}]$. Then
\[\Tr \big[(W_1\ldots W_{L/2})^{2}\big]=0,\quad
\mbox{if $(W_1\ldots W_{L/2})$ is fermionic}.\]
Exactly in this case we can commute one block around the trace and pick
up a sign.

Finally, we would like to determine the parity $P$ of an operator.
For an operator $\Tr (W_1\ldots W_L)$
to have a definite parity we need to have
\[\label{eq:parity}
(W_1\ldots W_L)=(W_i W_{i-1}\ldots W_2W_1W_L W_{L-1}\ldots W_{i+1})
\]
If this holds, parity is given by
\[P=(-1)^{[n/2]+[n'/2]+L}\]
where $n$ and $n'$ are the numbers of fermions in
$W_1\ldots W_i$ and $W_{i+1}\ldots W_L$, respectively.
This number is obtained by reversing the order of
$n+n'$ fermions and commuting $n$ fermions across $n'$.
Furthermore, each field has negative intrinsic parity, thus
the factor of $(-1)^L$.
If \eqref{eq:parity} does not hold, only the two combinations of states
\[
\Tr (W_1\ldots W_L)\pm \Tr (W_L\ldots W_1)
\]
have define and opposite parity.
For counting purposes, we should therefore assign some
parity to $\Tr (W_1\ldots W_L)$ and the opposite one
to the reversed operator $\Tr (W_L\ldots W_1)$.
We assign positive parity, $P=+$, to $\Tr (W_1\ldots W_L)$ if
\[
(W_1\ldots W_L)<(W_i W_{i-1}\ldots W_2W_1W_L W_{L-1}\ldots W_{i+1})
\quad \mbox{for all }i,
\]
and negative otherwise. In this way each pair of operators
receives both parities.



\end{document}